\documentclass{ar2e}
\usepackage{ulem}  % For Copy Editors
\usepackage{longtable} 
\usepackage{amsmath} 
                                                                             
\newcommand{\tab}{\hspace{5mm}}
\newcommand{\SU }{\ensuremath{SU(3)}\,\,}
\newcommand{\Tc}{\ensuremath{\theta_{C}} }
\input epsf
\newcommand{\cb}{\mbox{$\Xi^{0} \rightarrow
 \Sigma^{+}\, e^{-}\, \overline{\nu}_{e}\,\,$}}
\newcommand{\clpi}{\mbox{$\Xi^{0} \rightarrow \Lambda\, \pi^{0}$}\,}
\newcommand{\lppi} {\mbox{$ \Lambda  \rightarrow p\, \pi^{-}  $}}     
\newcommand{\lb}{\mbox{$\Lambda \rightarrow p e^{-}\overline{\nu}_e  $}\,}
\newcommand{\csp}{\mbox{$\Xi^{0} \rightarrow \Sigma^{+} \, \pi^{-}\,$}}
\newcommand{\fisq}{\mbox{$ f_{1}^{2}$}}
\newcommand{\gisq}{\mbox{$ g_{1}^{2}$}}
\newcommand{\fjsq}{\mbox{$ f_{2}^{2}$}}
\newcommand{\gjsq}{\mbox{$ g_{2}^{2}$}}
\newcommand{\fifj}{\mbox{$ f_{1} f_{2}$}}
\newcommand{\gigj}{\mbox{$ g_{1} g_{2}$}}
\newcommand{\figi}{\mbox{$ f_{1} g_{1}$}}
\newcommand{\figj}{\mbox{$ f_{1} g_{2}$}}
\newcommand{\fjgi}{\mbox{$ f_{2} g_{1}$}}
\newcommand{\fjgj}{\mbox{$ f_{2} g_{2}$}}

\newcommand{\sppi}{\mbox{$\Sigma^{+}  \rightarrow p\, \pi^{0} $}}
\newcommand{\snenu}{\mbox{$\Sigma^{-}  \rightarrow n\,e^{-}\overline{\nu}\,\,$}}
\newcommand{\spls}{\mbox{$\Sigma^{+}\,$}}
\newcommand{\lam}{\mbox{$\Lambda\,$}}

\newcommand{\suf}{\mbox{$ SU(3)\,$}}
\newcommand{\piz}{\mbox{$\pi^{0}\,$}}

\newcommand{\gf}{\mbox {$ g_1 / f_1\,$}}
\newcommand{\gfans}{\mbox {$1.32 \pm^{0.21}_{0.17} ( \mathrm{stat} ) \pm 0.05 ( syst ) $}}
\newcommand{\gfansst}{\mbox {$1.32 \pm^{0.21}_{0.17} ( \mathrm{stat} ) $}}
\newcommand{\scf}{\mbox {$ g_2 / f_1\,$}}

\newcommand{\wf}{\mbox {$ f_2 / f_1\,$}}
\newcommand{\wfans}{\mbox {$2.0 \pm 1.2 ( \mathrm{stat} ) \pm 0.5 ( \mathrm{syst} ) $}}
\newcommand{\xpe}{\mbox {$ x_{pe} $}}

\newcommand{\xpntr}{\mbox {$ x_{p\nu\perp} $}}
\newcommand{\xentr}{\mbox {$ x_{e\nu\perp} $}}
\newcommand{\cas}{\mbox{$\Xi^{0}\,$}}

\newcommand{\asyp}{\mbox{$\alpha_{\Xi} \alpha_{\Lambda} $}}
\begin{document}

%MTWinEqns1Page 24\tab 12/9/01\\

\title{Semileptonic Hyperon Decays}

\author{Nicola Cabibbo
\affiliation{Department of Physics, University of Rome - La Sapienza\\
Piazzale A. Moro 5, 00185 Rome, Italy\\
nicola.cabibbo@roma1.infn.it}
Earl C. Swallow
\affiliation{Department of Physics, Elmhurst College\\ Elmhurst, Illinois 60126\\ 
and\\ 
Enrico Fermi Institute\\ 
The University of Chicago\\ Chicago, Illinois 60637\\
earls@elmhurst.edu}
Roland Winston
\affiliation{Department of Physics and Enrico Fermi Institute\\
The University of Chicago\\ 
Chicago, Illinois 60637\\
winston@uchicago.edu}}

\begin{abstract}
We review the status of hyperon semileptonic decays. The central issue is
the $V_{us}$ element of the CKM matrix, where we obtain 
$V_{us}=0.2250 (27)$. This value is of similar precision, but higher, than the one derived from $K_{l3}$, and in better agreement with the unitarity 
requirement,  $|V_{ud}|^2+|V_{us}|^2+|V_{ub}|^2=1$. We find that the Cabibbo model gives an excellent fit of the existing form factor data on baryon beta decays
($\chi^{2} = 2.96$ for 3 degrees of freedom) with $F + D = 1.2670 \,\pm 0.0030$,  $F - D = -0.341 \pm 0.016$, and no indication of flavour \suf breaking effects. We indicate the need of more experimental and theoretical work, both on
hyperon beta decays and on $K_{l3}$ decays.
\end{abstract}

\maketitle

\section{Introduction }

It is nearly forty years since Cabibbo proposed a model \cite{cab} for weak hadronic currents 
based on \SU  symmetry. This model led to detailed predictions for the beta decays of the 
baryon octet, in particular for the beta decays of hyperons.  It has taken nearly 
four decades to test this model experimentally. 
The model is now embedded in the standard model of quarks and leptons 
and their interactions, and its principles are most conveniently illustrated in 
terms of the quark triplet $u,d,s$. 
%%% (Fig.1) I do not have this figure or the next. Can we drop it?
The $d\rightarrow u, \,s\rightarrow u$ and lepton matrix 
elements of the weak current 
are in the ratio of $\cos{\Tc}:\sin{\Tc}:1$. The Lorentz structure of the current 
is $V-A$, vector-minus-axial-vector, and \Tc --- the Cabibbo angle --- is a parameter to be determined from experimental data. 

The generalization of Cabibbo universality to three generations of quarks
was given by Kobyashi and Maksawa \cite{km} with a weak current
with the flavour structure $\sum_{i k} \bar u_i V_{i k} d_k$, where 
$u_i =\{ u\, c\, t\}$ are the $Q=2e/3$ quarks, and 
$d_k =\{d\,s\,b\}$ the $Q=-e/3$ quarks. They observed that this extension could
accomodate $CP$ violation. The $3\times 3$ matrix $V$ is known
as the Cabibbo-Kobyashi-Maksawa (CKM) matrix, and $V_{ud}\approx \cos \Tc,
V_{us}\approx \sin\Tc$.

Quark beta decay is not accessible to 
experiment and we must rely on the next best available cases, the meson octet or the baryon octet. 
%%% (Fig.2) I do not have this figure. Can we drop it?
In this article we will concentrate on the study of the beta decays
in the baryon octet, which include  both strangeness-conserving decays and strangeness-changing decays. Among the first we find ordinary nuclear beta 
decay, in particular the beta decay
of free neutrons, but also the $\Sigma\rightarrow\Lambda$ beta decays.  Strangeness
 violating decays  include
$\Lambda\rightarrow p, \;\Sigma^{-}\rightarrow n,\; \Xi^{-}\rightarrow \Lambda,\; \Xi^{-}\rightarrow \Sigma^{0}$ and $\Xi^{0}\rightarrow \Sigma^{+}$ beta decays.

If one neglects \SU  breaking effects,  the ensemble of  
baryon beta decays is described by three parameters, \Tc and the 
$F$ and $D$ parameters for the axial-vector matrix elements. The second parameter arises because there are \emph{two} 
irreducible matrix elements for an octet current between two octets. 
It was left to experiments to find how well the data from all decays in the baryon octet are 
described by these two parameters. As is frequently the case in testing a fundamental 
theory, the experimental tests proved to be highly non-trivial. 
One encounters  both theoretical and experimental difficulties. 

On the experimental side, the beta-decay branching ratios of  hyperons  
are typically 1/1000 or less, requiring considerable skill and 
resources to separate the beta decays from dominant two-body 
decay backgrounds. The notable exception is the neutral cascade 
$\Xi^{0}\rightarrow \Sigma^{+}\, e^{-}\,\bar\nu$ decay,
which is easily distinguished from
$\Xi^{0}\rightarrow\Lambda\pi^{0}$, the only two body decay that is 
energetically allowed.

If one seeks a precision test of the theory, one has to disentangle the effects of 
forbidden contributions to hyperon decays, such as the weak magnetism form 
factor,  from those arising from the allowed contributions.
Considerable skill and resources, in this case polarized hyperon beams, are required to disentangle the individual form factor contributions. Thus, from the lowest lying hyperon 
(lambda) to the highest (cascade), the experiments have, at each 
stage, challenged state-of-the-art-techniques. 

Only now, with the recent measurement of $\Xi^0$ beta decay 
\cite{cb, ffcb} can we form a perspective of how well the model 
actually accounts
 for the data. As we shall see, it accounts for them extremely well.

On the theoretical side the main impediment to a model-independent
test of the theory is the lack of  a model-independent computation of effects 
arising from the breaking of \SU  symmetry --- in modern terms,  from the 
mass difference between the $s$ and the $d$ quarks. Although present 
data on baryon beta-decay are in excellent agreement with the predictions 
of unbroken $SU(3)$, a model independent evaluation
of $SU(3)$-breaking effects would play a key role in a precision test of the 
unitarity of the CKM matrix. The means for such a test are probably available
with the methods of  lattice QCD simulations that have seen remarkable 
advances in  recent years. 

The weak mixing parameters $V_{us}, V_{ds}$ are already today the 
best known entries in the CKM matrix, but an improvement would be very  valuable, 
as it can lead to a better check on the unitarity of the $V$ matrix.
The value of $V_{us}$ presented in the Particle Data Group \cite{PDG2002} is essentially
derived from $K_{l3}$ decays, while results from hyperon beta decays are 
given an ancillary role.  We are convinced that already today the value of  
$V_{us}$ obtained from hyperon decay is of comparable precision to the  
$K_{l3}$ one.
Further theoretical and experimental work should in the near future
improve the situation in both hyperon and $K_{l3}$ beta decays.

\subsection{Historical Note}

Of course, this is the picture from an early 21$^{st}$ century 
perspective. 
It is \textit{not} the way pieces of the puzzle unfolded. In 
recounting 
the history, one might start with L. B. Okun's rapporteur talk \cite{OkunRochester58}
at the 1958 ``Rochester Conference'' at CERN.  Here 
mesons and hyperons are represented as composite states; 
of pions, nucleons and lambdas (the Sakata model, \cite{SakataModel}). 
In fact, there is even the suggestion, later dropped, 
that lambda and cascade hyperons might be the constituents! 
Universality between the vector coupling of nuclear and mu meson 
beta-decay is incorporated; It is recognized that the reduced 
strength of $\Delta S=1$ 
hyperon beta decays is compatible with the decay rate  of $K_{l3}$
decays, hinting at a universal suppression of strangeness-changing 
leptonic decays. 

A key clue can be found in the thesis of Feynman's 
student, Sam  Berman  \cite{Berman}. The issue 
was the comparison between the experimental value of the Fermi 
coupling constant as deduced from neutron and muon beta decay: could  
the slight discrepancy be accounted for by radiative corrections?
Berman and Feynman discovered that radiative corrections for beta decay
had the opposite sign from that needed to close the gap
between beta decay and muon decay. 
The conclusion of the thesis was ``The 
disagreement between experiment and theory appears to be outside 
the limit of experimental error and might be regarded as an 
indication of the lack of universality even by the strangeness conserving 
part of the vector interaction.'' 

The closest encounter with the concept of a mixing angle 
appeared in a footnote to the PCAC paper by M. Gell-Mann and 
M. L\'{e}vy, \cite{Gell-Mann:np}, which comments on the possible interpretations of 
Berman's discrepancy: 
\begin{quote}
``Should this discrepancy be real, it would probably indicate a partial or total failure of of the conserved vector current idea. It might also mean, however,  that the current is conserved 
but with $G/G_V <1$. Such a situation is  consistent with universality if we consider the 
vector current for $\Delta S=0$ and $\Delta S=1$ together to be something like:
\end{quote} 
\begin{equation*}
 GV_\alpha^{(\Delta S = 0)} + GV_\alpha^{(\Delta S = 1)} = 
G_\mu \bar p \gamma_\alpha (n + \epsilon \Lambda)(1+\epsilon^2)^{-1/2} + ...
\end{equation*}
\begin{quote}
And likewise for the axial-vector current. If $(1+\epsilon^2)^{-1/2}=0.97$, then 
$\epsilon^2=.06$, which is of the right
order of magnitude for explaining the low rate of $\beta$ decay of the $\Lambda$
particle. There is, of course, a renormalization factor for that decay, so we cannot be sure 
that the low rate really fits in with such a picture.''
\end{quote} 
On reading this footnote, two questions arise: What are the missing terms
in the vector current, represented by ellipsis points? Why did Gell-Mann fail to 
elaborate this footnote into a complete solution after discovering the \SU  symmetry?
The most probable answer is that up to 1962 there was evidence for
the presence of $\Delta S = - \Delta Q$  weak transitions which could not 
easily fit in either the Sakata model or the octet model. 

One of the first published examples of a hyperon  beta decay event was a
$\Delta S = - \Delta Q$ event, $\Sigma^{+}\rightarrow n \, 
\mu^{+} \nu$ observed  in emulsions \cite{AngelaBG}.
This event appeared  in a total observed sample of 
$120\, \Sigma^{+}$  decays. When in 1963 the first ``large statistics''
studies of hyperon decays were emerging at CERN, and no new 
$\Delta S = - \Delta Q$ event appeared,  a rough evaluation of the 
relevant branching ratios  convinced Cabibbo that the evidence for
for $\Delta S = - \Delta Q$ could be disregarded. This observation was crucial in working out
the consequences of  \SU  for the weak interactions, because by neglecting the 
possibility of $\Delta S = - \Delta Q$ one could adopt the simplest hypothesis,
according to which the weak current was a member of an \SU  octet.
Several years later  this event and other evidence for $\Delta S = - \Delta Q$  
decays was  dismissed as background \cite{Cronin68}. 

The present PDG limit \cite{PDG2002},  based on fairly old (pre-1975) experiments, 
\begin{equation*}
\label{sqlim}
\frac{\Sigma^+\rightarrow n \, \ell^+ \nu}{\Sigma^-\rightarrow n \, \ell^- \bar \nu}
< 0.043
\end{equation*}
could and should be substantially improved.

In 1963 Cabibbo \cite{cab} proposed 
a theory of the weak current, parameterized by a single mixing angle 
\Tc, in the context of the octet model of  \SU  symmetry.
The central assumption was that the weak current $J_\alpha$ is a member of an
octet of currents $J^i_\alpha = V^i_\alpha  + A^i_\alpha $, where $V^i_\alpha$ 
and $A^i_\alpha$ are octets of vector and axial currents,
\begin{equation}
\label{CabibboJ}
	J_\alpha=\cos\Tc (J^1_\alpha+iJ^2_\alpha)+
			\sin \Tc (J^4_\alpha+iJ^5_\alpha)
\end{equation}
By assuming that the vector 
and axial parts of the weak current are ``parallel,'' i.e. the same element of the respective
octets, the theory included the $V-A$ hypothesis\footnote{
Instruments for a comparison of the relative strength of the axial and vector 
currents were offered by Gell-Mann's current algebra \cite{GellMannCA}. 
The relevant test was executed with the Adler-Weisberger sum rule  \cite{AdlWeisbrg}.
}, and it also included the 
Conserved Vector Current (CVC ) hypothesis,  by assuming that the vector part 
of the weak current belongs to the same octet as the electromagnetic current.  

The theory led to a very detailed description of semileptonic decays 
of baryons and
mesons in terms of a small number of parameters, leading to
predictions for the matrix elements that have proven durable 
and remarkably accurate.

When expressed in terms of quarks, which were only
proposed in 1964,  the weak current of Ref.  \cite{cab} takes the simple
form
\begin{equation}
\label{CabibboQ}
	J_\alpha=\cos\Tc \,\bar u \gamma_\alpha (1+\gamma_5) d+
			\sin \Tc \,\bar u \gamma_\alpha (1+\gamma_5) s
\end{equation}

\subsection{Intimations of $CP$ violation}

Soon after Cabibbo proposed the quark mixing hypothesis, the 
suggestion was made \cite{gla_cab} that the same picture could 
naturally accommodate $CP$ violation by allowing the mixing angle 
to be complex (adding a phase). However, it was easily seen that 
a $2\times2$ unitary matrix (in the case of four quarks) can always be reduced 
to a form with real elements, and thus necessarily preserves $CP$. 

In 1973  Kobayashi and Maskawa noted that the mixing of three quark
families entails a single complex phase that cannot be eliminated
by field redefinitions. They thus proposed that the four-quark model should be
extended to a six-quark model in which mixing offers a natural 
explanation for the existence of $CP $ violation. Their proposal predated
by four years the discovery of $\Upsilon$ particles, the first experimental
detection of a fifth quark, the $b$ quark.

In the standard model with six quarks  the network of transition amplitudes 
between the charge --1/3  quarks, $d,s,b$,  and the charge 2/3 quarks, 
$u,c,t$ is described
by a unitary matrix $V$, the  CKM matrix, whose effects can be seen as a mixing
 between the $d,s,b$ quarks, 

\begin{equation}
	\label{CKM}
	\begin{pmatrix} 
		d^\prime\\ s^\prime\\ b^\prime 
	\end{pmatrix} =
	\begin{pmatrix}
		V_{ud} &  V_{us} & V_{ub}  \\
 		V_{cd}&  V_{cs}&V_{cb}  \\
 		V_{td}&  V_{ts}&V_{tb}  
	\end{pmatrix}
	\begin{pmatrix} d\\ s\\ b \end{pmatrix}
\end{equation}
The new description  can be seen as an extension of the current mixing
of Eq. \eqref{CabibboJ}, or the quark mixing in Eq.  \eqref{CabibboQ},
the relation between the two formulations being given by
\begin{equation}
\label{theta_CKM }
	\tan\Tc = \frac{V_{us}}{V_{ud}}.
\end{equation}

The recent observation of \emph{direct} $CP$ violation 
in the neutral kaon system \cite{KTeVepsi, Lai:2001ki} and the  $CP$ 
violation in $B$-meson oscillations 
\cite{Babar, Belle} are in brilliant confirmation of this picture
\cite{CabibboLP01}. 

\subsection{Status in 1983-4}

Hyperon beta decay has an important role in the study of 
weak interactions, both by establishing the 
validity of the predicted pattern of branching ratios and form 
factors, and by contributing to the determination of the parameter 
\ensuremath{\sin\Tc}. The first task is substantially achieved at the 
present level of  accuracy, and has been completed with the 
recent results on the beta decay of the neutral cascade. The road to the 
present satisfactory state of  affairs was not easy, and we will mention 
some of the difficulties which were encountered on the way.

At the time of the last \textit{Annual Review of Nuclear and Particle 
Science} article on hyperon decays, by Gaillard and Sauvage, 
\cite{Gaillard:1984ny}  nearly twenty years ago, the 
key experimental tests of the Cabibbo theory had not yet been 
done. While it is true that lambda beta decay had been found 
to be approximately $V-A$ as required, the theory faced its first 
critical test in sigma minus beta decay. Taking the $F/D$ ratio 
from any two decays, say neutron and lambda, the theory required 
sigma minus beta decay to appear $V+A$. This surprising sign reversal 
was considered at the time the ``go or no-go'' test of the Cabibbo 
theory. Measuring the sign convincingly required polarized sigma 
minus. Low energy experiments relied on tertiary polarized sigmas 
from pions or kaons. The statistics were meager, the control 
of systematics problematic, and the results less than compelling. If 
anything, the $V-A$ sign appeared favored by the data \cite{Keller}.

The turning point was determined by an experimental innovation,
the invention and development of hyperon beams. When such beams 
were  discovered to be significantly polarized, a 
new era of precision experiments with excellent control of 
systematics was inaugurated. With the ability to produce \textit {and 
reverse} polarization, correlations between momenta and polarization could 
be measured in a precise and bias-cancelled way. New precise experiments
settled the question of the $\Sigma^-$ beta decay in favor of the Cabibbo 
prediction (see Section 3.2).  The high-energy hyperon beam proved to be the 
enabling technology 
for carrying out precision measurements of hyperon decay properties.

\section{Theoretical Issues}

In this section we discuss different issues that must be taken into account
in an accurate treatment of hyperon semileptonic decays. These include 
the issue of the breaking of flavour  \SU  symmetry, radiative corrections, and the 
$q^2$ dependence of the form factors. We will however start from a discussion of the general form of the matrix element, which we will express in terms of a 
convenient 2-component spinor formulation \cite{bri}, and of the different observables in hyperon decays.

\subsection{ Baryon Matrix Elements }
The  $V - A$ transition matrix element for the generic 
hyperon beta-decay process $B \rightarrow \,b \,e^-\, \bar\nu$, where
$B$ and $b$ are the initial and final-state baryons,
can be written in the form 
\begin{equation}
\label{}
\mathcal{M}=\frac{G_S}{\sqrt{2}}
		\bar u_b\, (O^V_\alpha+O^A_\alpha)\, u_B\, 
		 \bar u_e \, \gamma^\alpha\, (1+\gamma_5)\,  v_\nu
\end{equation}
where 
\begin{align}
\label{}
     O^V_\alpha&= f_1(q^2) \gamma^\alpha +\frac{f_2(q^2)}{M_B}\sigma_{\alpha\beta}q^\beta +
				\frac{f_3(q^2)}{M_B} q_\alpha\\
     O^A_\alpha&=\left(g_1(q^2) \gamma^\alpha +\frac{g_2(q^2)}{M_B}\sigma_{\alpha\beta}q^\beta +
				\frac{g_3(q^2)}{M_B} q_\alpha\right)\gamma_5
\end{align}
The momentum transfer is $q^\alpha = (p_e + p_\nu )^\alpha = 
(p_B - p_b )^\alpha $ and the coupling strength
$ G_S=G_F V_{us}$ for $|\Delta S| = 1$ and 
$ G_S=G_F V_{ud}$ for $\Delta S = 0$, where $ G_F$ is the Fermi coupling 
constant, and $V_{us}, V_{ud}$ are the appropriate CKM 
matrix elements. We employ the metric 
and $\gamma$-matrix conventions of Ref. \cite{GarciaBook}\footnote{
These conventions are essentially those of Bjorken and Drell \cite{BjDrell},  
with two 
exceptions; $\gamma_5$ is defined with an opposite sign, and $\sigma_{\alpha\beta}
= \frac{1}{2}[\gamma_\alpha\, , \, \gamma_\beta]$ is defined without an $i$.}.

The vector and axial part of the weak current are members of two octets,
\begin{align}
\label{VAOct}
     V^i_\alpha&  = \bar q\frac{\lambda^i}{2}  \gamma_\alpha q,\nonumber\\
     A^i_\alpha&  = \bar q\frac{\lambda^i}{2} \gamma_\alpha \gamma_5 q,
\end{align}
where $\lambda^i/2$ are generators of \SU. Neglecting  the mass difference between the $s$ and the $u,\, d$ quarks, 
an explicit breaking of flavor   \SU  symmetry,  the form factors for baryon beta 
decays are related to each other. 
Matrix elements of an  \SU octet operator $O_k$ between octet states can in fact be expressed in terms of two reduced matrix elements, $F_O,\,D_O$ 
\begin{equation}
\label{OctetLaw}
\langle B_n| \,O_k\, | B_m\rangle  = F_O\,f_{knm} +D_O\, d_{knm}
\end{equation}
where $f_{knm}$  are the structure constants of \SU and $d_{knm}$
are defined by the anticommutation relations $\{\lambda^k,\lambda^n\}=
2\delta_{kn}+2d_{knm}$.  If we know the value of two 
independent matrix elements in Eq. \eqref{OctetLaw}, all of them can be 
determined. 

The vector part of the weak current and the 
electromagnetic current belong to the same octet, so that  the matrix elements  
of the  weak current can be predicted on the basis
of the electromagnetic form factors of the proton and neutron. This is true both of 
$f_1$ and of the ``weak magnetism'' form factor, $f_2$.

The allowed contribution of the axial current derives from the  $g_1$ form factor.
The leading contribution is  described by the two parameters, $F$ and $D$.  We will return to the problem  of the $q^2$ dependence of  $g_1$. 

The ``weak electricity'' form factor,  $g_2$,  vanishes in the limit of exact $SU(3)$,
as can be seen from a very simple argument:
flavor  \SU  symmetry connects  the axial weak current to two neutral currents,  
$A^3_\alpha$ and $A^8_\alpha$. Since the latter are even under charge 
conjugation, their matrix elements cannot contain a $g_2$ term, which could 
only arise from an odd-$\mathcal C$ current, a ``second-class current'' according 
to the terminology of S. Weinberg \cite{WeinbergII}.  It is easily seen that second 
class currents are excluded in the framework of the standard model.

Summing up, the expression \eqref{CabibboQ} of the  weak current
 leads, in the limit of exact flavor  \SU  symmetry, to simple 
predictions for the complex of baryon beta decays, in terms of 
\Tc  and  two further parameters, $D,F$, needed to describe the axial-vector 
form factors in the various decays. The $D,F$ parameters are the 
generalization to  \SU of the reduced matrix element of the 
Wigner-Eckart theorem for $SU(2)$. The vector form factor $f_{1}(0)$ 
is directly predicted in terms of \Tc and the nucleon electric charges.  The weak 
magnetism form factor $f_{2}(0)$ is predicted in terms of \Tc and 
the well-measured 
values of neutron and proton magnetic moments. The predictions for 
all of the octet baryons are given in Table \ref{Table:su3}.

% This is the SU(3) Cabibbo Model prediction Table
\begin{table}[htb]
\begin{center}
\caption{Cabibbo model predictions for octet baryon beta decays }
\begin{tabular}{@{}lcccccc@{}}
\hline
\hline
Decay   &  Scale  &  $f_1(0)$    &  $g_1(0)$ &  $g_1/f_1$ &  $f_2/f_1$  \\  
\hline
$n \rightarrow p e^- \overline{\nu}$ & $V_{ud}$  &  1  &  $D+F$ &  $F+D$  & $\frac{M_n}{M_p} \frac{(\mu_p - \mu_n)}{2}$ = 1.855   \\
$\Xi^- \rightarrow \Xi^0 e^- \overline{\nu}$ &  $V_{ud}$  &  -1  &  $D-F$ &  $F-D$  & $\frac{M_{\Xi^-}}{M_p} \frac{(\mu_p + 2\mu_n)}{2}$ = -1.432   \\
$\Sigma^\pm  \rightarrow \Lambda e^\pm  \nu$ &  $V_{ud}$  &  0$^a$  &    $\sqrt(\frac{2}{3})D$ & $\sqrt(\frac{2}{3})D$  &  $-\frac{M_{\Sigma^\pm}}{M_p} \sqrt(\frac{3}{2}) \frac{(\mu_n)}{2}$ = 1.490   \\
$\Sigma^- \rightarrow \Sigma^0 e^- \overline{\nu}$ &  $V_{ud}$  &  $\sqrt(2)$  &    $\sqrt(2)F$ &  $F$  & $\frac{M_{\Sigma^-}}{M_p} \frac{(2\mu_p + \mu_n)}{4}$ = 0.534   \\
$\Sigma^0 \rightarrow \Sigma^+ e^- \overline{\nu}$ &  $V_{ud}$  &  $\sqrt(2)$  &  $-\sqrt(2)F$  &  $-F$  & $\frac{M_{\Sigma^0}}{M_p} \frac{(2\mu_p + \mu_n)}{4}$ = 0.531   \\
$\Xi^0 \rightarrow \Sigma^+ e^- \overline{\nu}$ &  $V_{us}$  &  1  &  $D+F$  &  $F+D$  & $\frac{M_{\Xi^0}}{M_p} \frac{(\mu_p - \mu_n)}{2}$ = 2.597   \\
$\Xi^- \rightarrow \Sigma^0 e^- \overline{\nu}$ &  $V_{us}$  &  $\frac{1}{\sqrt(2)}$  &  $\frac{1}{\sqrt(2)}(D+F)$  &  $F+D$   & $\frac{M_{\Xi^-}}{M_p} \frac{(\mu_p - \mu_n)}{2}$ = 2.609   \\
$\Sigma^- \rightarrow n e^- \overline{\nu}$ &  $V_{us}$  &  -1  &  $D-F$  &  $F-D$  & $\frac{M_{\Sigma^-}}{M_p} \frac{(\mu_p + 2\mu_n)}{2}$ = -1.297   \\
$\Sigma^0 \rightarrow p e^- \overline{\nu}$ &  $V_{us}$  &  $\frac{-1}{\sqrt(2)}$  &  $\frac{1}{\sqrt(2)}(D-F)$  &  $F-D$  & $\frac{M_{\Sigma^0}}{M_p} \frac{(\mu_p + 2\mu_n)}{2}$ = -1.292   \\
$\Lambda \rightarrow p e^- \overline{\nu}$ &  $V_{us}$  &  -$\sqrt(\frac{3}{2})$  &  $-\frac{1}{\sqrt(6)}(D+3F)$  &  $F+\frac{D}{3}$  & $\frac{M_{\Lambda}}{M_p} \frac{\mu_p}{2}$ = 1.066   \\
$\Xi^- \rightarrow \Lambda e^- \overline{\nu}$ &  $V_{us}$  &  $\sqrt(\frac{3}{2})$  &  $-\frac{1}{\sqrt(6)}(D-3F)$ &  $F-\frac{D}{3}$  & $-\frac{M_{\Xi^-}}{M_p} \frac{(\mu_p + \mu_n)}{2}$ = 0.085   \\
\hline
\multicolumn{6}{l}{$^a$Since $f_1(0)=0$ for $\Sigma^\pm \rightarrow \Lambda e^\pm  \nu$, predictions are given for $f_2$ and $g_1$ rather than } \\
\multicolumn{6}{l}{$f_2/f_1$  and  $g_1/f_1$. } \\
\multicolumn{6}{l}{Here $\mu_p = 1.7928$, $\mu_n = -1.9130$, and $g_2 = 0$ for all decays.  } \\
\end{tabular}
\label{Table:su3}
\end{center}
\end{table}

\subsection{ The Effective Hamiltonian}
The beta decay matrix element for hyperon decay can be displayed in a form that 
is particularly accessible for analysis of experiments. This 
is because there are two small parameters; $q/M$, where $q$ is the 
momentum transfer and $M$ a baryon mass, and $m_{e}/M$, where $m_{e}$  is 
the electron mass. Retaining terms up to second order in $q/M$ 
is sufficiently accurate for analysis of current data as well 
as for the foreseeable future. Terms of order $m_{e}/M$ can be safely neglected
for hyperon decays, so that we can omit the scalar and pseudoscalar form factors
$f_3, g_3$. Hyperon muonic decays, where these form factors
are relevant,  have very small branching ratios and
can be omitted from any precision study of the quark mixing 
parameters. The matrix 
element can be written in an effective two-component form, Ref.  \cite{bri}, a 
technique first introduced by Primakoff \cite{pri1,pri2} in the context of muon 
capture.  We can then define an effective  Hamiltonian for the decay 
$B\rightarrow b e \bar \nu$, so that
\begin{equation}
\label{Minv}
{\mathcal M } =  \langle be \mid {\mathcal H_{\mbox{eff}}} \mid B \nu \rangle 
\end{equation}
with
\begin{eqnarray}
\label{Heff}
{\mathcal H_{\mbox{eff}}} & = & \sqrt{2}
G_{S} \: \frac{1 - {\vec{\mathbf{\sigma}_{\ell}}} \cdot \hat{e}}{2} \,
\bigg [ G_{V} + G_{A} {\vec{\mathbf{\sigma}_{\ell}}} 
\cdot  {\vec{\mathbf{\sigma}_{b}}} \nonumber \\
&  & + G_{P}^{e}  {\vec{\mathbf{\sigma}_{b}}}\cdot \hat{e} 
+ G_{P}^{\nu}  {\vec{\mathbf{\sigma}_{b}}} \cdot \hat{\nu}\bigg ] \,
\frac{1 - {\vec{\mathbf{\sigma}_{\ell}} }\cdot \hat{\nu}}{2}.
\end{eqnarray}
Here $\hat{e}$ and $\hat{\nu}$ are unit vectors along the
electron and antineutrino directions, while $e$, $\nu$, and $E_{B} $ are
the energies of the electron, antineutrino,
and initial baryon.
The spin operators ${\vec {\mathbf{\sigma}_{\ell}} }$ and 
$  {\vec{\mathbf{\sigma}_{b}}}$ act respectively on the two-component 
lepton and final state baryon spinors.

The effective coupling 
coefficients in Equation 11, $G_{V}$, $ G_{A}$, $G_{P}^{e}$, and $G_{P}^{\nu}$
are functions of the form factors, but depend also on the
frame of reference. The rest frames of the initial baryon $B$ and of
the final baryon $b$ are of particular interest in analyzing a beta decay experiment. The rest frame of the initial baryon is 
particularly convenient in analyzing the decay of polarized baryons,
and in this frame we have\footnote{
In the following we will use expressions which are correct to second order in the 
small parameter $q/M \, \approx\, \delta = (M_B-M_b)/M_B$.}
\begin{eqnarray}
G_{V} & =  & f_{1} - \delta f_{2} - \frac{\nu+e}{2M_{B}}
(  f_{1} + \Delta f_{2} ),  \nonumber \\
G_{A} & = & -g_{1} + \delta g_{2} + \frac{\nu-e}{2M_{B}}
(  f_{1} + \Delta f_{2} ),  \nonumber \\
G_{P}^{e} & = & \frac{e}{2M_{B}}\left[ - ( f_{1} + \Delta f_{2} )
 + g_{1} + \Delta g_{2} \right],  \nonumber \\
G_{P}^{\nu} & = & \frac{\nu}{2M_{B}}( f_{1} + \Delta f_{2} 
 + g_{1} + \Delta g_{2} ),  
\label{Bpol}
\end{eqnarray}

In the rest frame of the final-state baryon,   particularly convenient
if its polarization is an observable, the effective couplings are given by
\begin{eqnarray}
G_{V} & =  & f_{1} + \delta f_{2} - \frac{\nu+e}{2M_{B}}
(  f_{1} + \Delta f_{2} ),  \nonumber \\
G_{A} & = & -g_{1} + \delta g_{2} + \frac{\nu-e}{2M_{B}}
(  f_{1} + \Delta f_{2} ),  \nonumber \\
G_{P}^{e} & = & \frac{e}{2M_{B}}\left[ - ( f_{1} + \Delta f_{2} )
 - g_{1} + \Delta g_{2} \right],  \nonumber \\
G_{P}^{\nu} & = & \frac{\nu}{2M_{B}}( f_{1} + \Delta f_{2} 
 - g_{1} + \Delta g_{2} ),  
\label{Bfpol}
\end{eqnarray}
In either case $ \delta = (  M_{B} -   M_{b} ) / M_{B} $ 
and $ \Delta = (  M_{B} +   M_{b} ) / M_{B} = 2 - \delta $.

\subsection{ Decay Distributions}
We give the distributions to second order in the parameter 
\ensuremath{\delta}, 
which is sufficient for the accuracy of experiments. Of course, 
to this order one must also take the $q^2$ dependence of the form factors 
into account, as will be discussed later. As noted above, we will
distinguish two cases:\begin{enumerate}
   \item The energy spectrum and angular distribution of the 
	leptons is studied with respect to the polarization of the initial baryon.
	This case will be analyzed in the rest frame of the decaying baryon, with 
	the effective form factors given by Eq. \eqref{Bpol}.
   \item The energy spectrum and angular distribution of the 
	leptons is studied together with the polarization of the final state baryon.
	This case will be analyzed in the rest frame of the emitted baryon, with 
	the effective form factors given by Eq. \eqref{Bfpol}.
\end{enumerate}

Let us start with the first case, where we work in the rest frame of the decaying 
baryon, which has a polarization $\mathbf P_B$, and we do not measure the 
polarization of the final state particles.

The differential decay rate is
\begin{equation}
\label{dGamma}
d\Gamma = \frac{|\mathcal M|^2}
{(2\pi)^5}\frac{E_b+M_b}{2M_B}\frac{e^2\nu^3}{(e^{\textrm{max}}-e)}
		de\, d\Omega_e\, d\Omega_\nu
\end{equation}
After summing over final spins and averaging over the initial spin, $|M|^2$ is given by:
\begin{eqnarray}
\mid {\mathcal M } \mid^{2} &  = & G_{S}^{2}\,\xi \,\,
\big [1+a \hat{e} \cdot \hat{\nu} +
\textsf{A} \mathbf{P}_B \cdot \hat{e} + \textsf{B} \mathbf{P}_B \cdot \hat{\nu}  \nonumber \\
& & 
+ \textsf{A}' ( \mathbf{P}_B \cdot \hat{e} ) ( \hat{e} \cdot \hat{\nu} ) 
 + \textsf{B}' ( \mathbf{P}_B \cdot \hat{\nu} ) ( \hat{e} \cdot \hat{\nu} )  
+ \textsf{D} \mathbf{P}_B \cdot ( \hat{e} \times  \hat{\nu} ) \big ]
\end{eqnarray}
where $ \mathbf{P}_B$ is the polarization vector of the decaying baryon,  
and the different coefficients can be expressed in terms of the form factors 
of Eq. \eqref {Bpol} according to
\begin{align}
\xi = & \mid\!G_{V}\!\mid^{2} + 3  \mid\!G_{A}\!\mid^{2} 
 - 2 \textrm{Re}\left[ G_{A}^{*}( G_{P}^{e} +  G_{P}^{\nu} )\right]%\nonumber \\
 %& 
+ \mid\!G_{P}^{e}\!\mid^{2}
+ \mid\!G_{P}^{\nu}\!\mid^{2}, 
\nonumber \\
\xi a = & \mid\!G_{V}\!\mid^{2} - \mid\!G_{A}\!\mid^{2} 
- 2 \textrm{Re}\left[ G_{A}^{*}( G_{P}^{e} +  G_{P}^{\nu} )\right]\nonumber \\
&
+ \mid\!G_{P}^{e}\!\mid^{2}
+ \mid\!G_{P}^{\nu}\!\mid^{2} 
+ 2 \textrm{Re}(  G_{P}^{e*} G_{P}^{\nu} )( 1
 + \hat{e} \cdot \hat{\nu} ), 
\nonumber \\
\xi \textsf{A} = & - 2 \textrm{Re}(  G_{V}^{*} G_{A} ) - 2  \mid\!G_{A}\!\mid^{2} 
  + 2 \textrm{Re}(  G_{V}^{*} G_{P}^{e} +  G_{A}^{*} G_{P}^{\nu} ), \nonumber \\
\xi \textsf{B} = & - 2 \textrm{Re}(  G_{V}^{*} G_{A} ) + 2  \mid\!G_{A}\!\mid^{2}  
+ 2 \textrm{Re}(  G_{V}^{*} G_{P}^{\nu} - G_{A}^{*} G_{P}^{e} ), \nonumber \\
\xi \textsf{A}'  = & 2 \textrm{Re}(  G_{P}^{e*} (  G_{V} + G_{A} ) ), \nonumber \\
\xi \textsf{B}' = & 2 \textrm{Re}(  G_{P}^{\nu*} (  G_{V} - G_{A} ) ), \nonumber \\
\xi \textsf{D}   = & 2 \textrm{Im}(   G_{V}^{*} G_{A} ) + 2 \textrm{Im}(   G_{P}^{e} G_{P}^{\nu*} ) 
( 1 +  \hat{e} \cdot \hat{\nu} ) 
  + 2 \textrm{Im}\left[ G_{A}^{*} (  G_{P}^{\nu} -  G_{P}^{e} )\right]. 
\end{align}

We next consider the case where the energy spectrum and angular 
distribution of the leptons is 
studied with respect to the polarization of the final-state baryon,
$\mathbf P_b$.
Electron and antineutrino spins are again not observed; however, 
this case focuses on measurement of the final baryon  
polarization.  The invariant matrix element is in this case given by
\begin{eqnarray}
\mid {\mathcal M } \mid^{2} &  = & G_{S}^{2}\,\xi \,\,
\big [1+a \hat{e} \cdot \hat{\nu} +
\textsf{A} \mathbf{P}_b \cdot \hat{e} +  \textsf{B} \mathbf{P}_b \cdot \hat{\nu}  \nonumber \\
& & 
+ \textsf{A}' ( \mathbf{P}_b \cdot \hat{e} ) ( \hat{e} \cdot \hat{\nu} ) 
 + \textsf{B}' ( \mathbf{P}_b \cdot \hat{\nu} ) ( \hat{e} \cdot \hat{\nu} )  
%\nonumber \\
%& &
+ \textsf{D} \mathbf{P}_b \cdot ( \hat{e} \times  \hat{\nu} ) \big ]
\end{eqnarray}
and the different coefficients are given by
\begin{align}
\xi = & \mid\!G_{V}\!\mid^{2} + 3  \mid\!G_{A}\!\mid^{2} 
 - 2 \textrm{Re}\left[ G_{A}^{*}( G_{P}^{e} +  G_{P}^{\nu} )\right]
 %& 
+ \mid\!G_{P}^{e}\!\mid^{2}
+ \mid\!G_{P}^{\nu}\!\mid^{2}, 
\nonumber \\
\xi a = & \mid\!G_{V}\!\mid^{2} - \mid\!G_{A}\!\mid^{2} 
- 2 \textrm{Re}\left[ G_{A}^{*}( G_{P}^{e} +  G_{P}^{\nu} )\right]\nonumber \\
&
+ \mid\!G_{P}^{e}\!\mid^{2}
+ \mid\!G_{P}^{\nu}\!\mid^{2} 
+ 2 \textrm{Re}(  G_{P}^{e*} G_{P}^{\nu} )( 1
 + \hat{e} \cdot \hat{\nu} ), 
\nonumber \\
\xi \textsf{A} = & - 2 \textrm{Re}(  G_{V}^{*} G_{A} ) + 2  \mid\!G_{A}\!\mid^{2} 
  + 2 \textrm{Re}(  G_{V}^{*} G_{P}^{e} -  G_{A}^{*} G_{P}^{\nu} ), \nonumber \\
\xi \textsf{B} = & - 2 \textrm{Re}(  G_{V}^{*} G_{A} ) - 2  \mid\!G_{A}\!\mid^{2}  
+ 2 \textrm{Re}(  G_{V}^{*} G_{P}^{\nu} + G_{A}^{*} G_{P}^{e} ), \nonumber \\
\xi \textsf{A}'  = & 2 \textrm{Re}(  G_{P}^{e*} (  G_{V} - G_{A} ) ), \nonumber \\
\xi \textsf{B}' = & 2 \textrm{Re}(  G_{P}^{\nu*} (  G_{V} + G_{A} ) ), \nonumber \\
\xi \textsf{D}   = & 2 \textrm{Im}(   G_{V}^{*} G_{A} ) + 2 \textrm{Im}(   G_{P}^{e*} G_{P}^{\nu} ) 
( 1 +  \hat{e} \cdot \hat{\nu} ) 
  + 2 \textrm{Im}\left[G_{A}^{*} (  G_{P}^{e} -  G_{P}^{\nu} ) \right]. 
\end{align}
The polarization of the final baryon may be expressed 
explicitly as:
\begin{equation}
\mathbf{P}_b = \frac{ (\textsf{A} + \textsf{A}' \hat{e} \cdot \hat{\nu} ) \hat{e}
+ (\textsf{B} + \textsf{B}' \hat{e} \cdot \hat{\nu} ) \hat{\nu}
+ \textsf{D}  \hat{e} \times  \hat{\nu} }{ 1 + a \hat{e} \cdot \hat{\nu} }.
\end{equation}
The components of this polarization can
readily be measured when the outgoing 
baryon $b$ is a hyperon which undergoes a subsequent 
weak decay $ b  \rightarrow b' \pi $ with a nonzero decay 
asymmetry parameter $ \alpha_{b'} $.  The distribution of the $ b' $ 
direction relative to any axis defined by a unit 
vector $ \hat{i} $ is given by 
\begin{equation}
\frac{1}{\Gamma} \;  \frac{d \Gamma }{d \Omega_{b'}} =
\frac{1}{4 \pi }( 1 + {\mathsf{S}_{i}} \alpha_{b'} \,\hat{i} \cdot \hat{b'} ),
\end{equation}
where $ {\mathsf{S}_{i}} =  \langle {\mathbf{P}_{b}} \cdot \hat{i} \rangle $ 
is the average 
polarization of $b$ in the $ \hat{i}$ direction.
Conceptually, it is advantageous to employ the orthonormal basis
\begin{eqnarray}
\hat{\alpha} & = & 
\frac{  \hat{e} + \hat{\nu} }{ \sqrt{ 2(1 
+ \hat{e} \cdot \hat{\nu})}}, \nonumber \\
\hat{\beta}  & = & 
\frac{  \hat{e} - \hat{\nu} }{ \sqrt{ 2(1 
- \hat{e} \cdot \hat{\nu})}}, \nonumber \\
\hat{\gamma} & = &  \hat{\alpha} \times \hat{\beta}.
\end{eqnarray}
Experimentally, it may be more advantageous to
 determine the polarization components along
one or more of the outgoing particle
 directions ($ \hat{e},  \hat{\nu}, \hat{b} $).

Analytic expressions for the integrated final state polarization 
$\textsf{S}_{e}$, 
$\textsf{S}_{\nu}$,$\textsf{S}_\alpha$,$\textsf{S}_\beta$ 
are given in Appendix 
A up to (and including) second order in \ensuremath{\delta}. 
We observe 
that $\textsf{S}_\alpha$ depends solely on $V\times A$ terms, while 
$\textsf{S}_\beta$ depends solely on $V\times V$ and $A\times A$ 
terms in conformity with Weinberg's 
theorem \cite{Weinbergthm}. Another 
useful symmetry relation is that $\textsf{S}_{e}$ and $\textsf{S}_{\nu}$ are 
the same as those for a polarized initial baryon (hyperon) in the zero 
recoil (\ensuremath{\delta} \ensuremath{=} 0) limit. Stated another 
way, the lepton correlations with respect to $\mathbf  P_{B}$, the polarization of 
the \textit{initial state} baryon, are related to the correlations 
with respect to $\mathbf  P_{b}$, the polarization of the \textit{final state} 
baryon by interchanging $e$ and \ensuremath{\nu} throughout and reversing the 
sign of $\textsf{D}$.

\subsection{ $q^{2}$ Dependence of Form Factors}

To obtain expressions which are correct to $O(q^2)$, we can neglect
the $q^{2}$ dependence of the form factors $f_2,\,g_2$, whose contribution to
the transition amplitude is already $O(q)$.
In the limit of exact  \SU  symmetry, the $q^{2}$ dependence of
the vector form factor $f_1$ can be expressed 
\begin{eqnarray}
\label{fq2}
f^{bB}_1(q^2) &=& C^{bB}_F\,F_1(q^2) + C^{bB}_D\, D_1(q^2)\\
		 &=& C^{bB}_F[F_1(0) + \lambda_{F_1}  \,q^2] 
			+ C^{bB}_D\, [D_1(0) + \lambda_{D_1}  \,q^2] \nonumber
\end{eqnarray}
where $C^{bB}_F, C^{bB}_D$ are the appropriate $f$ and $d$ constants
(Eq. \ref{OctetLaw} ), and $F_1(q^2), D_1(q^2) $ the corresponding
reduced form factors, that can be expressed in terms of the measured  
charge form factors of the proton and neutron \cite{GarciaBook}, leading to
\begin{equation}
\label{fq2f1}
F_1(0)=1,\; D_1(0)=0,\; \lambda_{F}= \,6.13\, \mathrm{GeV}^{-2} ,\,
	 \lambda_{D} = \,0.12\, \mathrm{GeV}^{-2}
\end{equation}
The axial current
form factor, $g_1$,  can only be related to neutrino reactions, but the data
are not sufficient to determine two separate slopes for the $D$ and $F$ parts.
We thus follow Ref. \cite{Gaillard:1984ny} and use a dipole form,
e.g. in neutron beta decay case,
\begin{equation}
\label{}
g_1^{np}(q^2) = \frac{g_1^{np}(0)}{(1-q^2/M_A^2)^2}
\end{equation}
with $M_A = 1.08\pm 0.08$ $\mathrm{GeV/c^2}$.
A similar parametrization for the  vector form factor gives
\begin{equation}
\label{}
f_1^{np}(q^2) = \frac{f_1^{np}(0)}{(1-q^2/M_V^2)^2}
\end{equation}
with $M_V  = 0.84 \pm 0.04$ $\mathrm{GeV/c^2}$.  For the 
$\Delta S = 1$ decays, the scaling argument of Ref. \cite{Gaillard:1984ny} yields
$M_V  = 0.97$ $\mathrm{GeV/c^2}$ and $M_A = 1.25$ $\mathrm{GeV/c^2}$.  

\subsection{ Radiative corrections }
Radiative corrections to beta decay have been extensively studied 
by A. Sirlin \cite{Sirlin} and others \cite{radcor}. We have already 
noted the central role that these played, through Berman's work \cite{Berman}, 
in providing an important hint of universality breaking between 
neutron decay and muon decay. Hyperon decay experiments are only 
now becoming sufficiently sensitive to require radiative 
corrections, and then only for a limited set of observables. 
A standard reference is  \cite{GarciaBook}.

The situation can summarized as follows: 
\begin{itemize}
   \item Integrated observables 
	such as correlations with respect to initial or final baryon 
	polarization or the electron--neutrino correlation are practically 
	unaffected \cite{GarciaBook} to rather high accuracy, well beyond 
	the precision 
	of present and contemplated experiments. For these, radiative 
	corrections can safely be ignored. 
   \item Decay rates and the electron spectrum are affected and need to 
	be corrected before fitting for form factors. This applies, in 
	particular, to the weak magnetism form factor $f_2$ that is 
	sensitive to the electron spectrum. The total 
	rate (or branching ratio) is also affected to a few percent (for 
	example the total rate is increased by 4.4\% in $\Xi^0 \, \rightarrow\, 
	\Sigma^+ e^-\bar\nu_e$). Therefore 
	radiative corrections must be applied to the measured 
	beta-decay branching ratios for precise determinations of the 
	Cabibbo angle \Tc.
\end{itemize}

\subsection{  Flavor  \SU  breaking: The Ademollo-Gatto Theorem}
Symmetry breaking effects can be expanded
in powers of  $H'$,  the  $SU(3)$-breaking term in the 
hadron Hamiltonian. In the standard model, 
\begin{equation}
\label{Hprime}
H'  =  \frac{1}{\sqrt{3}} \left(m_s -\frac{m_d+m_u}{2}\right)\,\bar q \lambda^8 q
\end{equation}
In a previous section we have considered the expansion
of physical quantities  in  ``orders of forbiddenness,'' which has been
standard in the beta-decay literature since Fermi's 1934 paper.  
For a  hyperon decay $B \rightarrow b \,e^- \,\bar\nu$, this is equivalent
to an expansion in powers  of  $q/M_B$. 
 Here we are considering a similar expansion in powers of 
$H'$. Since $q \approx (M_B-M_b) \approx \, H' $,
the  two expansions can be combined in a single expansion
in powers of the small parameter $\delta = (M_B-M_b)/M_B$.
 For example, the first terms in the expansion of $f_1(q^2)$ (see \cite{GarciaBook}) are
\begin{equation}
\label{}
	f_1(q^2)  = C^{bB}_F + \Delta^1 f_1(0) +
			\Delta^2 f_1(0) + q^2 f_1^\prime (0) + \ldots,
\end{equation}
where $\Delta^1 f_1(0) ,\,\, \Delta^2 f_1(0)$ are respectively the first-order and
second-order $H'$ corrections, and the omitted terms are of 
third order or higher in $\delta$.  In the 
second-order expansion above we can use the exact-\SU result
in Eq. \ref{fq2}, \ref{fq2f1} for  $f_1^\prime (0)$, which
appears in a second-order correction ($\propto q^2$).

In 1964 Ademollo and Gatto proved \cite{ad} that  
there is no first-order correction to the vector form factor,
$\Delta^1 f_1(0)=0$. 
This is an important result: since experiments can measure 
$G_S f_1$,  knowing the value of  $f_1(0)$ in 
$\Delta S =1 $ decays is essential for determining $V_{us}$.
The theorem can be derived \cite{fufu,SakuraiAG}  from the 
commutation relations for the $\Delta S = 1$ weak vector charge,  $Q^{4+i5}$,
\begin{equation}
\label{crq}
\big [ Q^{4+i5},Q^{4-i5}  \big] = Q^3 + \sqrt{3} Q^8= Q^{\mathrm{em}}+Y
\end{equation}
For example, taking the expectation value in a baryon state $B=\Sigma^{-}$, we obtain the sum rule
\begin{eqnarray}
\label{crqe}
(Q^{\mathrm{em}}+Y)_B &=& - \big |\langle b|Q^{4+i5}|B\rangle  \big|^2\\
		&&+ \sum_m  \big |\langle m|Q^{4-i5}|B\rangle  \big|^2 
		- \sum_n \big|\langle n|Q^{4+i5}|B\rangle  \big|^2\nonumber
\end{eqnarray}
The states $|m\rangle , |n\rangle $ are hadronic states not in the baryon octet,
while $b$ belongs to the same  octet\footnote{
If $B=\Xi^-$,  two ``in octet'' states, $b=\Lambda, \Sigma^0$, can contribute.
This complicates the analysis, but the conclusions remain the same}
 as $B$. 
Following the Fubini-Furlan prescription, we work in the limit $P\rightarrow \infty$
as in this limit we find 
\begin{equation}
\label{pinfty}
      \lim_{P\rightarrow \infty}\langle b|Q^{4+i5}|B\rangle  = f_1(0)
\end{equation}
where $f_1$ is the vector form factor for the $B \rightarrow b \,e^- \,\bar\nu$
beta decay.   The matrix elements in the two sums vanish in the limit of
exact $SU(3)$, where $Q^{4+i5}$ transforms $B$ into members of the same octet,
and are thus $O(H')$, so that Eq. \eqref {crqe} leads to:
\begin{equation}
\label{crqe1}
f_1(0)^2 = - (Q^{\mathrm{em}}+Y)_B + O(H^{\prime 2})
\end{equation}
This proves the Ademollo-Gatto theorem: the first term on the r.h.s. represents the \SU pre\-diction, 
the second is the symmetry breaking correction,  and is indeed $O(H^{\prime 2})$.

This suggests a precise strategy in analyzing experiments to extract $V_{us}$. 
We use the information available from rates and angular correlations 
to extract in each decay the value of $V_{us} f_1(0)$. 
The Ademollo-Gatto theorem then guarantees that we can compute the value of 
$ f_1(0)$ with reduced sensitivity to symmetry breaking effects. Each decay thus 
provides a value for
$V_{us}$. If the theory is correct, these should coincide within errors,
and can be combined to obtain a best value of $V_{us}$.

\subsection{ Models for \SU breaking}

Treatments of \SU breaking effects fall essentially in two categories.
To the first belong those treatments which use group theory to determine the transformation 
properties of the correction, and accordingly introduce a number of parameters
to describe the pattern of deviations from \SU predictions in the different
decays. This strategy  was more attractive in the past --- when experimental data 
seemed to be in strong disagreement with the  theory ---  than it is today, when the experimental data are in excellent agreement with the ``exact \SU\!\!'' 
predictions of Ref. \cite{cab}.  

In the present situation, a fit that includes 
more parameters could at most be used to obtain upper limits on the deviation 
from the ``exact \SU\!\!'' case. This does not mean that deviations are absent, but only
that present data are not precise enough to establish their presence within
the ensemble of hyperon (and neutron) beta decays. 

If we wish to make progress in the understanding the deviations from exact 
flavour \SU 
we must resort to an explicit computation. Limiting our attention to
\SU\!\!-breaking corrections to the $f_1$ form factor, relevant for a determination 
of $V_{us}$ or $\sin\Tc$, we find in the literature computations that
use some version of the quark model, as in \cite{Donoghue:th,Schlumpf:1994fb},
or some version of chiral perturbation theory, as in \cite{Krause:xc, 
Anderson:1993as, Flores-Mendieta:1998ii}. 

A modern revisitation
of the quark-model computations will probably be feasible in the near future
with the technologies of lattice QCD.  The quark-model 
computations find that the $f_1$ form factors
for the different $\Delta S=1$ decays are  reduced by a factor, 
the same for all decays, given as $0.987$  in \cite{Donoghue:th},  and $0.975$ 
in \cite{Schlumpf:1994fb},  a decrease respectively of $1.3\%$ or $2.5\%$.
This is a very reasonable result, the decrease arising from the 
mismatch of the wave functions of baryons containing different numbers
of the heavier $s$ quarks. We would expect that the same result would be obtained
in \textit{quenched} lattice QCD, an approximation that 
consists in neglecting components in the wave function of the baryons 
with extra quark-antiquark pairs. This is known to be an excellent
approximation in low-energy hadron phenomenology see \cite{quenched}.

Multiquark effects can be included in  lattice  QCD by forsaking the 
quenched approximation for a  \textit{full} simulation. Alternatively,
given the very high computational cost of full simulations, one could 
resort to models in order to capture the major part of the multiquark
contributions. It is here that chiral models could play an important 
role, since one could arguably expect the largest part of the multiquark
contribution to arise from virtual $\pi,\, K,\,\eta$ states. 

Calculations of  $f_1$ in chiral perturbation theory  range from 
small negative corrections  in  \cite{Krause:xc} to larger
positive corrections in \cite{Anderson:1993as,quenched}.
Positive corrections in $f_1$ for \textit{all} hyperon  beta decays 
cannot be excluded,
but are certainly not expected in view of an argument \cite{Quinn:1968qy} 
according to which one expects a negative correction to $f_1$ at least in the 
$\Sigma^{-}\rightarrow n \, e^{-} \bar\nu$ case.
This result follows by considering the sum rule in Eq.  \eqref{crqe} 
for the case $B=\Sigma^-$. The states $m$ that contribute to the first sum 
have quantum numbers $S=-2 ,\; I = 3/2$; no resonant baryonic state
is known with these quantum numbers.
If we accept the hypothesis that the sums 
 in Eq.   \eqref{crqe} are dominated by resonant hadronic states, 
we can conclude that the first sum is smaller than the second, 
so that the correction to $f_1$ in $\Sigma^-$ beta decay 
should be  negative. 
We note that the argument of Ref.  \cite{Quinn:1968qy} 
applies as well to $K_{l 3}$ decays, and that the corrections
to these decays, computed with chiral perturbation theory,
are, as expected, negative. 

\subsection{ Weak Magnetism}

In the  \SU  symmetry limit, 
the value of  $f_2(0)$ for the different decays (see Table \ref{Table:su3})
is described by two parameters, $F_\mu,\,D_\mu$,
which are fixed in terms of the proton and neutron (anomalous) magnetic moments,
\begin{equation}
\label{F2}
 	F_\mu = (2\,\mu_P + \,\mu_N )/2,\;\; D_\mu = - 3\mu_N/2
\end{equation}
\tab In contrast to $f_1(0),  \,f_2(0)$ is \emph{not} protected by the 
Ademollo-Gatto theorem. However, 
this quantity has not been measured to sufficient precision to 
reveal  \SU  symmetry breaking effects. 

We note an ambiguity in expressing the \SU limit 
that clearly indicates the relevance of first-order symmetry breaking: 
should Eq. \eqref{OctetLaw} be applied to $f_2(0)$ or to 
$f_2(0)/M_B$?   Which of the two choices has smaller \SU
breaking corrections?  The second choice is traditionally preferred \cite{Sirlinf2, E715},
and is the one we adopt.  In combination with the fact that
the magnetic form factor is normalized with $M_p$, this gives rise to 
the $M_B/M_p$ factors in Table \ref{Table:su3}.

\subsection{ Weak Electricity}

In the absence of second class currents \cite{WeinbergII}
 the form factor $g_2$ can be seen to vanish in the 
 \SU  symmetry limit. The argument is very straightforward: the 
neutral currents $A^3_\alpha= \bar q \lambda^3 \gamma_\alpha \gamma_5 q$
and $A^8_\alpha= \bar q \lambda^8 \gamma_\alpha \gamma_5 q$
that belong to the same octet as the weak 
axial current are even under charge conjugation, so that their 
matrix elements cannot contain a weak -- electricity term, which is $C$-odd.
The vanishing of the weak electricity in the proton
and neutron matrix elements of  $A^3_\alpha,\, A^8_\alpha$ implies 
the vanishing of the $D$ and $F$ coefficients for $g_2(0)$, 
so that, in the \SU limit, the $g_2(0)$ form factor vanishes 
for any current in the octet. 

In hyperon decays  a nonvanishing $g_2(0)$ form factor can arise from the 
 breaking of \SU  symmetry. Theoretical estimates \cite{Holstein} 
indicate a value for $g_2(0)/g_1(0)$ in the $-0.2$ to 
$-0.5$ range.

 In determining the axial-vector form factor $g_1$ from the Dalitz Plot --- or, equivalently, the electron--neutrino correlation --- one is actually measuring  
$\tilde g_1$, a linear combination of $g_1$  and $g_2$  
($\tilde g_1 \approx g_1 - \delta g_2$   up to first order in 
$\delta = \Delta M/M$). This has already been noticed in past experiments 
and is well 
summarized in Gaillard and Sauvage \cite{Gaillard:1984ny}, Table 8. 
Therefore, in deriving $G_s^2f_1^2$ (hence $V_{us}$) from the beta decay rate, there is in fact a small sensitivity to  $g_2$. 
To first order,  the rate is proportional to   $G_s^2[f_1^2 + 
3 g_1^2 - 4 \delta\, g_1 g_2] \approx G_s^2[f_1^2 + 3 \tilde g_1^2 +2 \delta\, \tilde g_1 g_2]$.

Experiments that measure correlations with polarization --- in addition to the electron--neutrino 
correlation --- are sensitive to $g_2$. While the data are not yet sufficiently precise to yield good quantitative information, one may nevertheless look for trends. In polarized  $\Sigma^-\rightarrow n\,e^-\bar\nu$  \cite{E715} 
negative values of $g_2/f_1$ are clearly 
disfavored (a positive value is preferred by $1.5\sigma$). Since the same experiment unambiguously established that $g_1/f_1$  is negative one concludes that allowing for nonvanishing $g_2$  would increase the derived value of  $G_s^2f_1^2$. 
In polarized  $\Lambda \rightarrow p\,e^-\bar\nu$ the data favor \cite{Oehme} negative values of  $g_2/f_1$  (by about $2\sigma$).  In this decay, $g_1/f_1$  is positive so that again, allowing for the presence of nonvanishing $g_2$  would increase the derived value of $G_s^2f_1^2$. In either case, 
we may conclude that making the conventional assumption of 
neglecting the $g_2$ form factor tends to \textit{underestimate} the derived 
value of $V_{us}$. 
A more quantitative conclusion must await more precise experiments. 

%begin of casbrev.tex
\section{Experiments}

\subsection {Experimental data on Hyperon Decays}

The aim of experiments on hyperon beta decay is to derive values for the form factors
 that can be compared to theory. In considering how to extract form factors from the data, 
it is useful to summarize the kinds of observations required for each. 
We have seen that the induced scalar and pseudoscalar form factors $f_3$ and $g_3$ are not observable 
because of the smallness of the electron mass.  For the others, the situation is summarized in Table \ref{Table:measff}, which  shows the central role of polarization in measuring the form factors --- in particular their relative signs.  
It may be stated that critical tests of the theory have depended on information from either the initial or final baryon polarization.  
% This is the Measuring Form Factors TAble....
\begin{table}[htb]
\begin{center}
\caption[Measuring the form factors]{The contribution of different measurements
to the determination of the form factors in hyperon beta decay. The $e-\nu$
correlation is only sensitive to the \textit{magnitude} of $g_1/f_1$
and $g_2/f_1$.\\}
\begin{tabular}{@{}lcccc@{}}
\hline
\hline
Measured Quantity            &   $f_1$   &   $f_2/f_1$   &   $g_1/f_1$   &  $g_2/f_1$   \\  
\hline
Branching Fraction           &   $\surd$  &          &         &          \\   
Polarization                 &          &          &  $\surd$  &    $\surd$   \\  
e$\,\,\nu$ Correlation   &  $\surd$   &          &  $\surd$  &  $\surd$   \\  
Electron Spectrum            &          &   $\surd$  &         &          \\  
\hline
\end{tabular}
\label{Table:measff}
\end{center}
\end{table}

Experimental results through the year 2002 are summarized in 
Table \ref{Table:datasum}. Values are drawn from the 2002 edition of \textit{Review 
of Particle Physics} \cite{PDG2002} unless noted otherwise. We have included 
error scale factors \cite{PDG2002}, $S$,  which account for inconsistencies between measurements.
Decay rates are calculated by 
dividing the beta decay branching fraction by the particle lifetime. 
The highly precise value of g$_{1}$/f$_{1}$ for neutron beta decay is 
derived primarily from measurements of the electron asymmetry 
parameter measured in experiments with polarized neutrons. For 
the hyperon decays, the g$_{1}$/f$_{1}$ \textit{magnitudes} are 
determined 
primarily from measurements of the electron-neutrino correlation 
parameter (or, equivalently, the recoil baryon energy spectrum) 
while the signs are unambiguously determined from measurements 
involving hyperon polarization. 
% This is the Beta Decay Data Summary Table
%\tabcolsep {3 }
\begin{table}[htb]
\begin{center}
\caption{Summary of octet baryon beta decay data }
\begin{tabular}{@{}l@{  }c@{  }c@{}c@{}c@{}}
\hline
\hline
Decay          &   Lifetime   &   Branching   &   Rate$^f$         &   $g_1/f_1$ \\  
Process        &     (sec)    &    Fraction   & ($\mu sec^{-1}$) &      \\  
\hline
$n \rightarrow p e^- \overline{\nu}$ &  $885.7(8)$  &  1  &  $1.1291(10) 10^{-9}$  &  $1.2670(30)^a$ \\
$\Lambda \rightarrow p e^- \overline{\nu}$ &  $2.632(20) ^b 10^{-10}$ &  $0.832(14) 10^{-3}$  & $3.161(58)$  &  $0.718(15)$  \\
$\Sigma^- \rightarrow n e^- \overline{\nu}$ & $1.479(11) ^c 10^{-10}$  & $1.017(34)10^{-3}$  & $6.88(24)$  & $-0.340(17)$   \\
$\Sigma^- \rightarrow \Lambda e^- \overline{\nu}$ &  $1.479(11) ^c 10^{-10}$  & $0.0573(27) 10^{-3}$  & $0.387(18)$  & $f_1/g_1 = -0.01(10) ^d$   \\
$\Sigma^+ \rightarrow \Lambda e^+ \nu$ &  $0.8018(26) 10^{-10}$  & $0.020(5) 10^{-3}$  & $0.250(63)$  &    \\
$\Xi^- \rightarrow \Lambda e^- \overline{\nu}$ &  $1.639(15) 10^{-10}$  & $0.563(31)10^{-3}$  & $3.44(19)$  &  $0.25(5)$  \\
$\Xi^- \rightarrow \Sigma^0 e^- \overline{\nu}$ &  $1.639(15)10^{-10}$  & $0.087(17)10^{-3}$  & $0.53(10)$  &    \\
$\Xi^0 \rightarrow \Sigma^+ e^- \overline{\nu}$ &  $2.900(90)10^{-10}$  & $0.257(19)^e 10^{-3}$  & $0.876(71)$  & $1.32(+.22/-.18)$   \\
\hline
\multicolumn{5}{l}{  $^a$S = 1.6 \hspace{4em} $^b$ S = 1.6  \hspace{4em}    $^c$ S = 1.3  \hspace{4em} $^d$ S = 1.5  } \\
\multicolumn{5}{l}{$^e$Mean of two independent measurements \cite{cb, aahDPF99} by the KTeV Collaboration.} \\
\multicolumn{5}{l}{$^f$The rate is simply the branching fraction divided by the lifetime.} \\
\end{tabular}
\label{Table:datasum}
\end{center}
\end{table}

We will describe those experiments where sufficient information is available to extract the vector form-factor $f_1$, 
emphasizing results obtained since the excellent review of Gaillard and Sauvage
\cite{Gaillard:1984ny}.
The experiments may be classified in two groups according to the experimental techniques used for producing the hyperons. 
The first group uses hyperon beams, a technique pioneered at the Brookhaven AGS, the CERN PS, and for
polarized hyperons, Fermilab - see the excellent review by Lach and Pondrom \cite{Pondrom}.
The second uses high intensity neutral beams developed for precision kaon experiments and adapted with great success for studies of neutral hyperons. 

\subsection {\snenu}

It has long been recognized \cite{CabChil} that the prediction of a \textit{negative} 
sign for $g_1/f_1$ in $\Sigma^-$  beta decay --- in contrast  to the \textit{positive} sign observed in neutron beta decay and other strangeness-changing hyperon beta decays --- is a characteristic feature of the flavour \SU structure in the Cabibbo model \cite{cab}.  Thus the determination of this sign is a pivotal qualitative test of the model.

In the allowed (zero-recoil) approximation, experiments on unpolarized $\Sigma^-$   are sensitive only to $|g_1/f_1|$ through the electron-neutrino correlation parameter 
$\alpha_{e\nu}$  or equivalently through the recoil neutron spectrum.  As discussed in the review of Gaillard and Sauvage \cite{Gaillard:1984ny}, 
the early CERN hyperon beam experiment WA2 \cite{Bourquin:1983III}
 obtained $|g_1/f_1| = 0.34 \pm 0.05$ for \snenu, in agreement with previous results.

On the other hand, experiments on polarized $\Sigma$ hyperons are sensitive to the sign of $g_1/f_1$  through interference effects in the parity-violating spin 
asymmetry parameters $\alpha_{e}$, $\alpha_{\nu}$, and  $\alpha_{n}$ 
as shown in Table \ref{Table:g1f1sign}.  Four earlier low-energy experiments \cite{Keller} obtained a total sample of 352 events 
with a combined electron asymmetry parameter value of $\alpha_{e} = +0.26 
\pm 0.19$.  The Cabibbo sign is clearly not compatible with this value.  
This discrepancy (about 4.5 standard deviations \cite{Keller}) inspired considerable theoretical speculation \cite{GarciaBohm}.  

% This is the Sigma beta decay table....
\begin{table}[htb]
\begin{center}
\caption{Sign of $g_1/f_1$ and decay asymmetry parameters for $\Sigma^- \rightarrow
n e^- \overline{\nu}$}
\begin{tabular}{@{}lcccc@{}}
\hline
\hline
Decay          &   Asymmetries for   &   Asymmetries for   &   Low-energy   &   Fermilab E715 \\  
Parameter        &  $g_1/f_1 = +0.34$   & $g_1/f_1 = -0.34$  & Experiments &      \\  
\hline
$\alpha_{e \nu}$  &  0.345   &    0.345   &         ---          &   +0.364 $\pm$ 0.029     \\   
$\alpha_e$   &  0.391   &    -0.603    &  +0.26 $\pm$ 0.19   &   -0.519 $\pm$ 0.104     \\  
$\alpha_{\nu}$   &  0.603   &    -0.391    &        ---          &   -0.230 $\pm$ 0.061      \\  
$\alpha_n$     &  -0.685   &   +0.685   &          ---            &   +0.509 $\pm$ 0.102      \\  
\hline
\end{tabular}
\label{Table:g1f1sign}
\end{center}
\end{table}

In the absence of $\Sigma^- $ polarization, the CERN WA2 collaboration sought to determine \cite{Bourquin:1983III}
the sign of $g_1/f_1$ from the first-forbidden distortion of the electron spectrum.  
This analysis favored a negative sign by about 2.6 standard deviations.  However, the sensitivity of the electron spectrum to $g_1/f_1$ is quite small 
(the shape is dominated by phase space) and quite sensitive to experimental biases, radiative corrections, and the assumed value for
 the induced form factor $f_2$ (weak magnetism).  

Small sample sizes, substantial background levels, and limited polarization control were among 
the clear limitations of the low-energy experiments with polarized $\Sigma^- $.  The existence of appreciable hyperon polarization 
[first observed for neutral hyperons \cite{Bunce}]
in hyperon beams \cite{Pondrom} produced at nonforward production angles opened the door to a definitive experiment, Fermilab E715.    

\begin{figure}[htb]
\epsfxsize=15.0cm
\begin{center}
\epsfbox{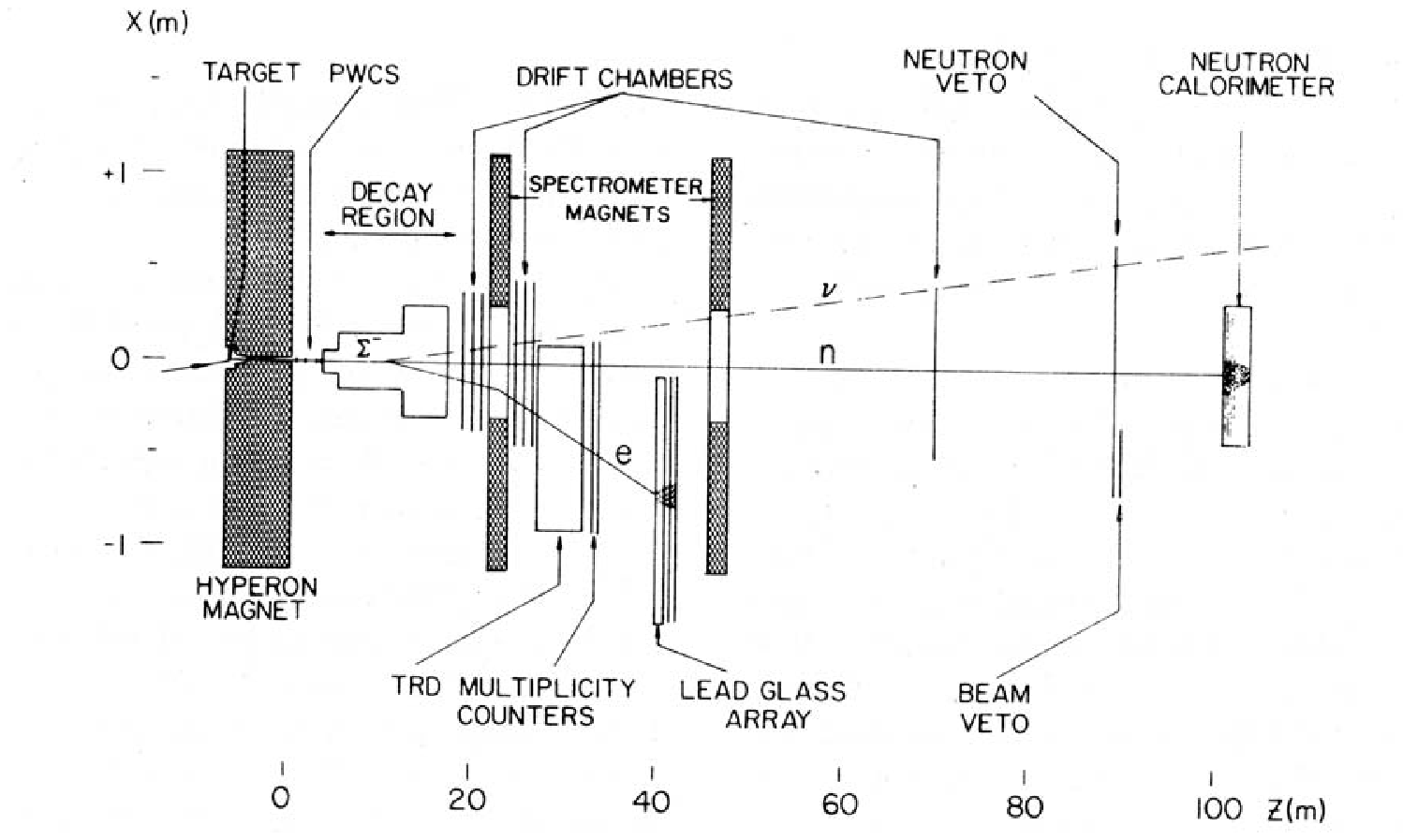}
\caption[The E 715 Detector ]{Plan view of the E715 apparatus, with typical particle trajectories. 
 The incident proton-beam angle corresponds to a positive targeting angle in the horizontal plane. 
 Note that the X and Z scales are different. }
\label{fig:E715}
\end{center}
\end{figure}

A plan view of the E715 experimental apparatus is shown in Fig. \ref{fig:E715}.
The experiment \cite{E715} was performed using the Fermilab Proton Center charged-hyperon beam. 
 Polarized hyperons were copiously produced at a nominal momentum of 250 $\mathrm{GeV}/\mathrm{c}$. 
 Changing the direction of the incident 400 $\mathrm{GeV}/\mathrm{c}$ 
proton beam readily altered the hyperon polarization direction. 
At an average production angle of 2.5 mrad, the measured $\Sigma^- $ polarization was $(23.6 \pm 4.3)\%$.  

To distinguish the relatively rare beta-decay mode, $\snenu$, from the dominant $\Sigma^- \rightarrow n\pi^-$ mode, the experiment employed 
double electron identification with a 12-plane transition radiation detector (TRD) and a four-layer lead-glass calorimeter array.  
High-pressure proportional chambers determined $\Sigma^- $ trajectories, and a drift-chamber magnetic spectrometer measured the momenta 
of charged decay products.  A neutron calorimeter located far downstream provided energy and direction measurements for the decay 
neutrons, allowing a full reconstruction of the beta decays.  
A sample of 49,671 candidate beta decays contained a background of less than 2\%.  

The ability to reverse the $\Sigma^- $ polarization (by alternating positive and negative targeting angles) made possible the use of bias-canceling 
techniques to determine $\alpha_{e}$, $\alpha_{\nu}$, and  $\alpha_{n}$ values as given in Table \ref{Table:g1f1sign}. In fact, some 
data were even recorded with the $\Sigma^- $ polarization
 perpendicular to the vertical hyperon magnet field, and the precession in the magnetic field used to determine the $\Sigma^- $ magnetic moment 
\cite{Zapalac} with both two-body decays and beta decays.   

The form factor ratios  $|g_1/f_1|$ and $f_2/f_1$ were determined most sensitively from the neutron and electron spectra respectively
 in the $\Sigma^- $ rest frame yielding $|g_1/f_1 -0.237g_2/f_1| = 0.327 \pm 0.020$ and $f_2/f_1 = -0.96 \pm 0.15$.  
A general fit that included the asymmetry parameters and made the conventional assumption $g_2 = 0$ 
gave the final value $g_1/f_1 = -0.328 \pm 0.019$.  As Table \ref{Table:g1f1sign} and the final value for $g_1/f_1$ show, this experiment unambiguously 
resolved the controversy concerning the sign of $g_1/f_1$ in favor of the Cabibbo model prediction.

\subsection {\lb }
(a) Brookhaven National Laboratory AGS experiment:
This landmark experiment \cite{WiseLambda} provided the first high statistics study 
of lambda beta decay. The neutral beam was derived at $4^o$ with respect to the primary proton beam, 
so that the lambdas were produced polarized. However, the polarization information was not used in the subsequent analysis. 
A sweeping magnet removed charged particles in the manner characteristic of hyperon beam arrangements, and a 10 radiation length 
lead filter removed photons leaving a neutral beam consisting mainly of lambdas, kaons and neutrons. 
The spectrometer consisted of two analyzing magnets and spark chambers to measure the laboratory momenta 
of the two charged decay particles with comparable precision. Four threshold Cerenkov counters were used for particle identification. 
The data sample after cuts consisted of slightly over 10,000 \lb and 25,000 \lppi events. 
This yielded a precision measurement of the branching ratio: 
BR = $  \lb/\lppi = (0.843 \pm 0.017)10^{-3}$.
Using the world average for the lifetime known at that time 
\cite{PDG78} the absolute rate for $\Lambda^0$ beta decay can be derived:
$\Gamma(\lb)=(3.204 \pm 0.068)\,10^{-3} s^{-1}$.
The form factor analysis was made on the basis of the Dalitz plot that reflects the electron neutrino angular correlation. 
Although the lambdas were produced \textit{in  principle} polarized, this information was not used because the  targeting angle was not reversed. 
This precluded making use of the bias canceling technique subsequently 
used to advantage by the \snenu  experiment \cite{JensenPrivate}.
Without  polarization correlations the form factor result is a linear combination of  $g_1$ and $g_2$ that the authors 
give as $|g_1/f_1|= 0.715+0.25 g_2/f_1$. In the final result, the  form factor 
$g_2$ is set to zero  and the final result given as $|g_1/f_1|= 0.715\pm0.026$.
With these provisos, the form factor results are: 
\begin{equation*}
f_1=1.238\pm 0.024,\; |g_1|=0.885\pm 0.030,\;f_2=1.34\pm 0.20.
\end{equation*}

The values stated are extrapolated to  $q^2=0$ and modified for radiative corrections. 
Although the sign of the form factors cannot be readily deduced from 
unpolarized data, the sign of $g_1$   can be safely set to be positive 
on the basis of earlier lower statistics experiments that measured 
correlations with lambda polarization \cite{Lindquist}.
The determination of  $|g_1/f_1|$ from the Dalitz plot in the Brookhaven experiment does not depend on the value of $f_2$.

(b) Fermilab neutral hyperon beam: This experiment \cite{dwo} is the highest statistics measurement of  \lb  to date, having analyzed
 nearly 40,000 events. In conception this experiment has many similarities
with the BNL one. A sweeping magnet removed charged particles, 
a characteristic feature of hyperon beams. Both a  threshold Cerenkov counters and a lead-glass array were used 
for particle identification. The spectrometer consisted of an analyzing magnet and multiwire proportional chambers to measure the 
laboratory momenta of the two charged decay particles. Again, the polarization information is not used in the data analysis. 
There is however a noteworthy innovation in the  event reconstruction. Typically,  the $\Lambda^0$ momentum (hence the neutrino momentum) is reconstructed with a two-fold ambiguity because the $\Lambda^0$ direction
 in the laboratory is well-known, but the energy is not. 
Thus, in fitting for  \lb  there is a two-fold ambiguity in 
the angle $\cos \theta^*_{e\nu}$ between the electron and anti-neutrino in the lambda rest frame, which diminishes its analyzing power. 
The authors point out that there is no ambiguity in the momentum of the proton-electron system considered as a fictitious particle, $\mathcal{Q}$. 
Thus the decay sequence is $\Lambda^0\rightarrow\mathcal{Q} +\bar\nu$  and  
$\mathcal{Q}\rightarrow p+e$. In the laboratory system, consider a plane P perpendicular to the direction of the neutral beam. By momentum conservation, the intersections of $\mathcal{Q}, p, e$  with this plane are collinear, as are the 
intersections of $\Lambda^0 , \mathcal{Q},\bar\nu$. Since there is no twofold kinematic ambiguity in these lines of intersection, the distribution of the 
 included angle is more sensitive to $|g_1/f_1|$  than the distribution in  
$\cos \theta^*_{e\nu}$. This method of analysis was used to advantage
 in a succeeding experiment on  \cb\!\!; we will discuss it more extensively
in the section dedicated to that experiment. 

The result for the form factors is:  $|g_1/f_1|= 0.731\pm 0.016$. 
While the result is given with positive sign on the basis of a slight sensitivity of induced terms to sign, we prefer to rely on experiments that
include polarization correlation, where the effects are large. The caveats with respect to $g_2$ dependence apply, 
and this form factor is set to zero as in the BNL experiment. The values stated are extrapolated to $q^2 =0$ and modified for radiative corrections.
 In addition, a value for the weak magnetism form-factor is given: $f_2/f_1= 
0.15\pm 0.30$. This is quite far from the expected value of $f_2/f_1\approx 1$. 
In fact, if one uses the expected value for $f_2$, the result changes slightly:  
$|g_1/f_1|= 0.719\pm 0.016\pm 0.012$. 

(c) CERN SPS Charged-Hyperon Beam: An experiment at the charged-hyperon beam at CERN collected over 7,000   \lb
events among a number of charged hyperon beta decay channels. \cite{Bourquin:1983I}. The experiment is well-described in  \cite{Gaillard:1984ny}. 
We summarize it here for completeness. The lambdas arise from $\Xi^-\rightarrow  \Lambda^0\pi^-$   decays, so that each $\Lambda^0$ is tagged. 
This gives the charged-hyperon beam an advantage over neutral beams where  $K^0$ decays, and in particular  $K^0\rightarrow \pi^+ e^-\bar\nu$,  
can be a significant source of background. Electron identification relied on both lead glass and transition radiation detectors to suppress 
the dominant 2-body decay mode of the $\Lambda^0$. The form-factor analysis made use of baryon kinetic energy, electron kinetic energy, 
Dalitz plot and electron-neutrino correlation. Moreover, the $\Lambda^0$ is polarized with an  asymmetry parameter of  $\alpha=-0.456\pm 0.014$
\cite{PDG2002}, so that $g_1$  can be determined both in magnitude and sign. The result is  $g_1/f_1= 0.70\pm 0.03$
with $g_2$  taken to be 0. Otherwise, the sensitivity to $g_2$    is stated as  
$\Delta g_1/\Delta g_2= 0.20$. The weak magnetism form-factor was derived 
from the electron spectrum to have the value $f_2/f_1= 1.32\pm 0.81$ . The values stated are extrapolated to $q^2 =0$ and modified for 
radiative corrections.

\subsection {\cb}

This  may appropriately be called ``the last hyperon beta decay.'' It is the last 
of observable beta decays of the octet to be measured. The experiment was long considered sufficiently 
problematic to be below the radar of most compilations. A notable exception 
was the original Cabibbo proposal \cite{cab}. Paradoxically, this is in some respects the most accessible beta decay. The \spls is a unique signature for the beta decay mode (the analog two-body mode is forbidden by energy conservation) so that event samples are remarkably free of 
two-body backgrounds that typically plague experiments. Moreover, the 
final-state \spls polarization is self-analyzing because of the large asymmetry of the decay \sppi \, ($\alpha = -0.98$). It is thus sufficient to study angular correlations with the proton. However, attaining sufficient event rate   required a new generation of hyperon experiments that 
combine the high-energy advantages of  hyperon beams with the high phase space acceptance of  neutral kaon beams. 
Beams with the desired properties arose in the context of recent 
precision studies of $CP$ violation in neutral kaon decays  \cite{e'ktev, Lai:2001ki}. In place of the narrow phase-space selection of a hyperon beam magnet, the new experiments use an identifying feature of either the production or decay process. In effect, the event is ``tagged.'' In the case of the decay \cb, only the beta decay mode has sufficient $Q$-value to produce a \spls\!\!.  Identified \clpi 
can be used as the primary source for the study of \lam decay.

The KTeV experiment (Fermilab E799) reported the first observation \cite{cb} 
of the decay \cb, followed by a measurement of the form factors
\cite {ffcb}.  
The KTeV neutral beam is produced 
by an 800 $\mathrm{GeV / c} $ proton
beam hitting a 30 $\mathrm{cm}$ BeO target at
an angle of  4.8 $\mathrm{mrad}$.  
The sweeping magnets used to remove charged particles from the neutral beam
also serve to precess the polarization of the 
\cas to the vertical direction.  The polarity of
the final sweeping magnet is regularly flipped
so as to have equal numbers of \cas 
polarized in opposite directions, making the
ensemble average of the \cas polarization negligible. The presence of
both a CsI electromagnetic calorimeter and a system of transition
radiation detectors (TRD) permitted double electron identification,
a useful feature in beta decay experiments.

\begin{figure}[htb]
\epsfxsize=14.0cm
\begin{center}
%s \epsfxsize=20.cm
\epsfbox{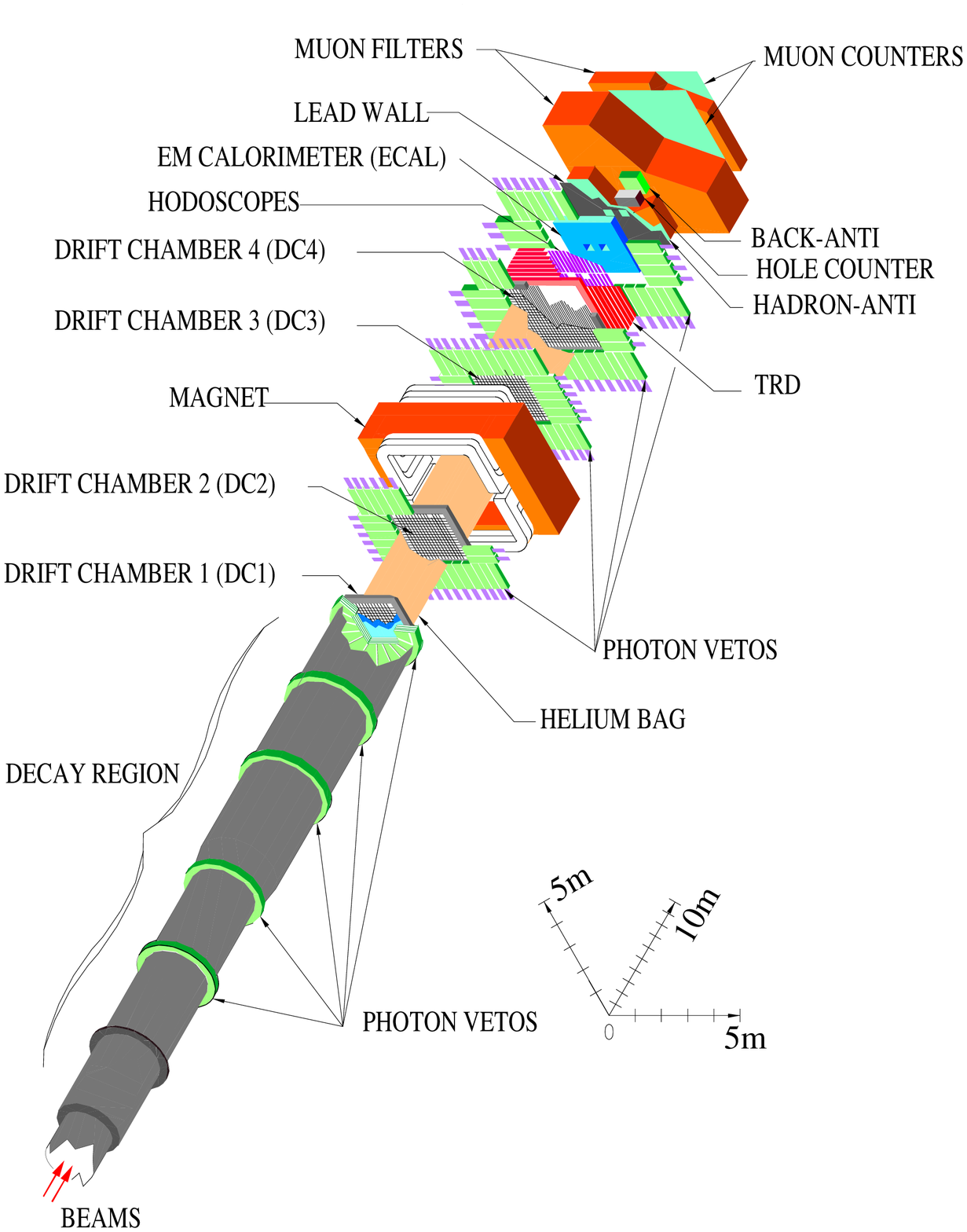}
\caption{ The KTeV Detector }
\end{center}
\end{figure}

\begin{figure}[htb]
\epsfxsize=8.0cm
\begin{center}
%s \epsfxsize=20.cm
\epsfbox{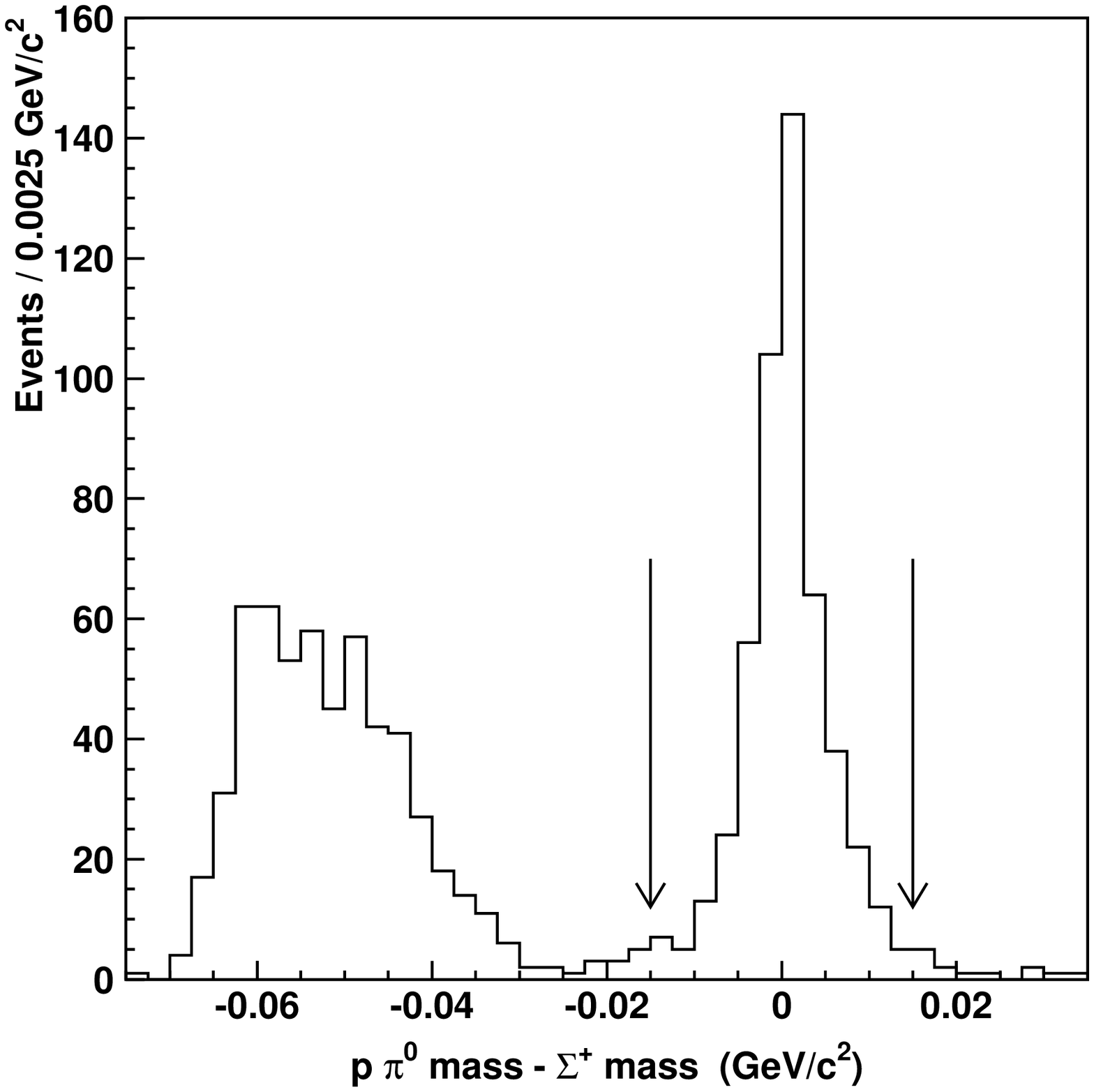}
\caption[The \sppi mass peak]{ The \sppi\, mass peak, after all selection 
criteria have been applied.
The background to the left of the peak is due
to \clpi decays ( followed by \lppi or \lb ).
Since \cb is the only source of \spls
in the beam (\csp is kinematically forbidden), 
signal events are identified by
having a $p$-\piz mass within 15 $\mathrm{MeV}$ 
of the nominal \spls mass.}
\label{fig:bkg}
\end{center}
\end{figure}

\begin{figure}[htb]
\epsfxsize=10.0cm
\begin{center}
%s \epsfxsize=20.cm
\epsfbox{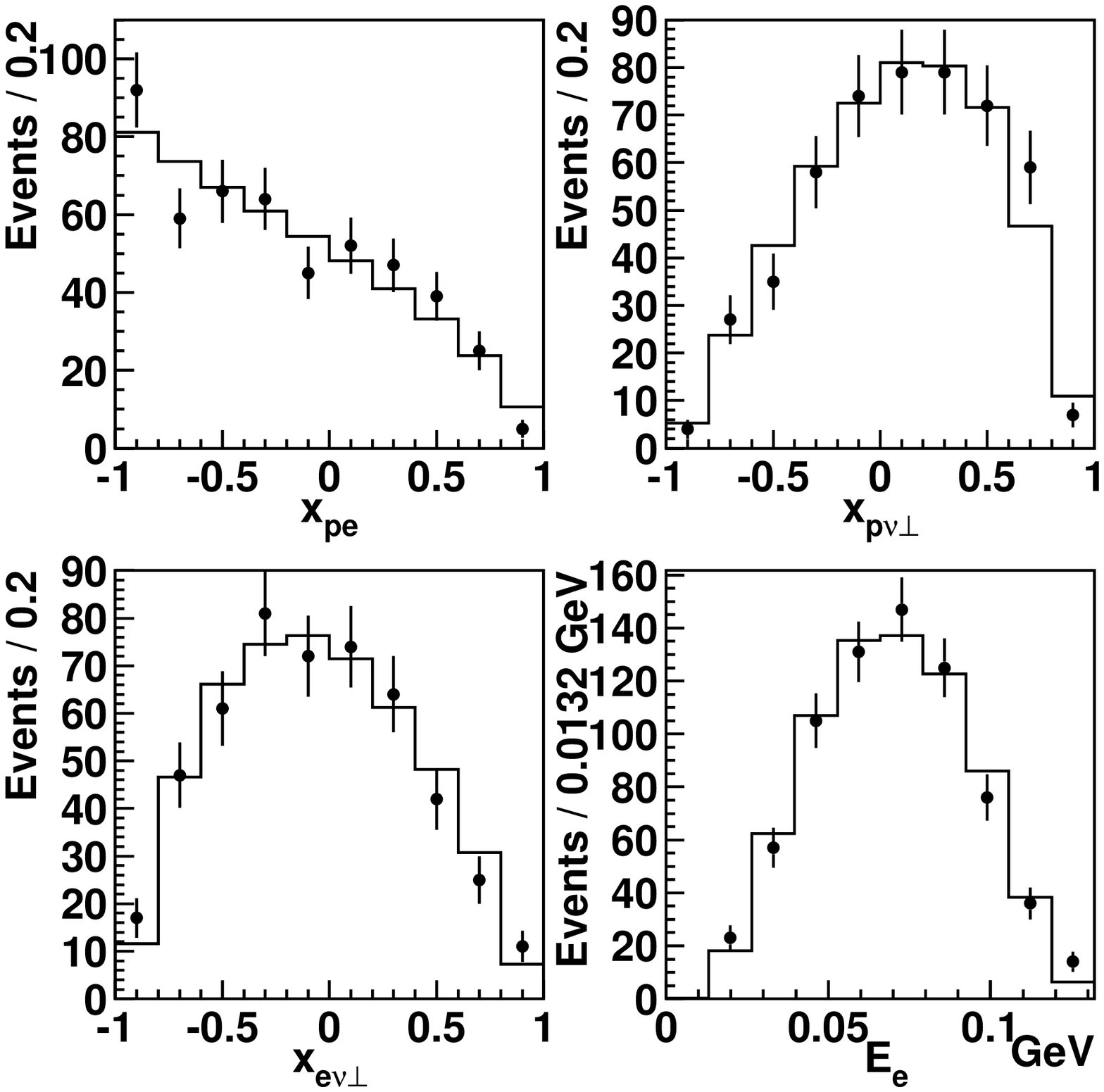}
\caption[The variables used to fit \gf and \scf]{ The three variables used to fit \gf and \scf,
and the energy spectrum of the electron in the \spls frame 
(used to determine \wf ).  The points
are data and the histogram is a Monte Carlo simulation with
\gf $ = 1.27 $ and $ \scf = 0. $ }
\label{fig:mk1d}
\end{center}
\end{figure}

Since the neutrino is unobserved, one cannot unambiguously reconstruct
the directions in the center of mass.  Instead one can obtain unambiguous 
angular variables transverse to the direction
of the \cas momentum\footnote{As discussed above, this method was first used in the Fermilab neutral 
hyperon beam experiment on \lam\, beta decay.}.  Following Dworkin 
 \cite{dwo}, we consider the decay sequence
\begin{equation}
\cas \rightarrow Q + \overline{\nu_{e}}, \,\,\,Q \rightarrow \spls + e^{-}
\end{equation}
where we have introduced the fictitious particle $Q$.
We then construct angular variables out of these
transverse quantities.  Denoting quantities in the $Q$
rest frame with an asterisk, we have the transverse momenta of the
electron, proton, and neutrino in the $Q$ frame:
${\vec{p}}_{e \perp}^{*}$, ${\vec{p}}_{p \perp}^{*}$ and
${\vec p}_{\nu \perp}$\footnote{ Since the $Q$ and the \cas
momenta are nearly parallel, ${\vec p}_{\nu \perp}$
is approximately equal to ${\vec p}_{\nu \perp}^{*}$.}.
The magnitudes of the momenta in the $Q$ frame are calculated
to obtain the  unambiguous kinematic quantities
\begin{equation}
\xentr  =  \frac{{\vec{p}}_{e \perp}^{*} \cdot {\vec{p}}_{\nu \perp}}
{{E_{e}}^{*} {E_{\nu}}^{*}} 
\end{equation}
and
\begin{equation}
\xpntr  =  \frac{{\vec{p}}_{p \perp}^{*} \cdot {\vec{p}}_{\nu \perp}}
{\mid  {\vec{p}}_{p}^{*} \mid {E_{\nu}}^{*}} 
\end{equation}
which correspond to the polarization of the \spls along the
neutrino direction, and the electron-neutrino correlations
respectively.  The kinematic quantity corresponding 
to the proton-electron correlation
is \xpe, the cosine of the angle between the proton and the
electron in the \spls frame.  The one dimensional
distributions for \xpe, \xentr \,and \xpntr \,are shown in Fig. \ref{fig:mk1d}.

\begin{figure}[tttt]
\epsfxsize=8.0cm
\begin{center}
%s\epsfxsize=20.cm
\epsfbox{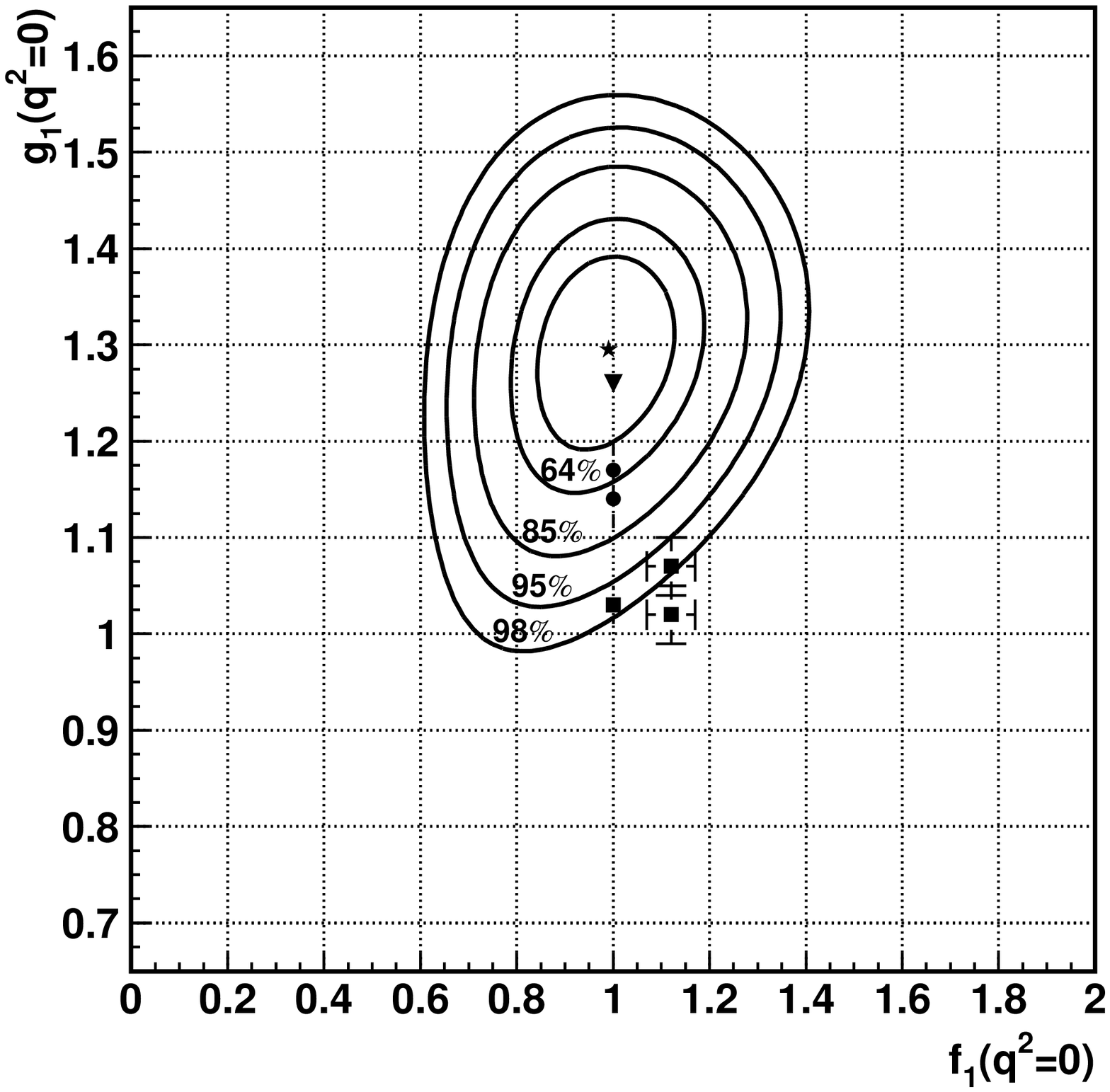}
\caption[Confidence interval plot for $f_1$ and $g_1$]{Confidence interval plot for $f_1$ and $g_1$.
The inverted triangle is exact SU(3) symmetry; the star indicates the KTeV value. Solid circles and squares are SU(3) breaking fits
from \cite{rat}and \cite{Flores-Mendieta:1998ii} respectively.} 
\label{fig:gf}
\end{center}
\end{figure}

To determine \gf\!, a maximum likelihood fit
for \gf using \xpe, \xpntr \, and \xentr\, is performed. After correcting 
for background, the final value for \gf is \gfansst (Fig. \ref{fig:gf}).
As a check of the Monte Carlo simulation,
the two body asymmetry product \asyp 
is determined with a sample of $70,000$ \clpi events.
The measured \asyp $= -0.286 \pm 0.008 (\mathrm{stat}) \pm 0.015 (\mathrm{syst})$
is consistent with the Particle Data Group 
value of $ -0.264 \pm 0.013 $ \cite{PDG2002}.

\begin{figure}[htp]
\epsfxsize=8.0cm
\begin{center}
%s \epsfxsize=20.cm
\epsfbox{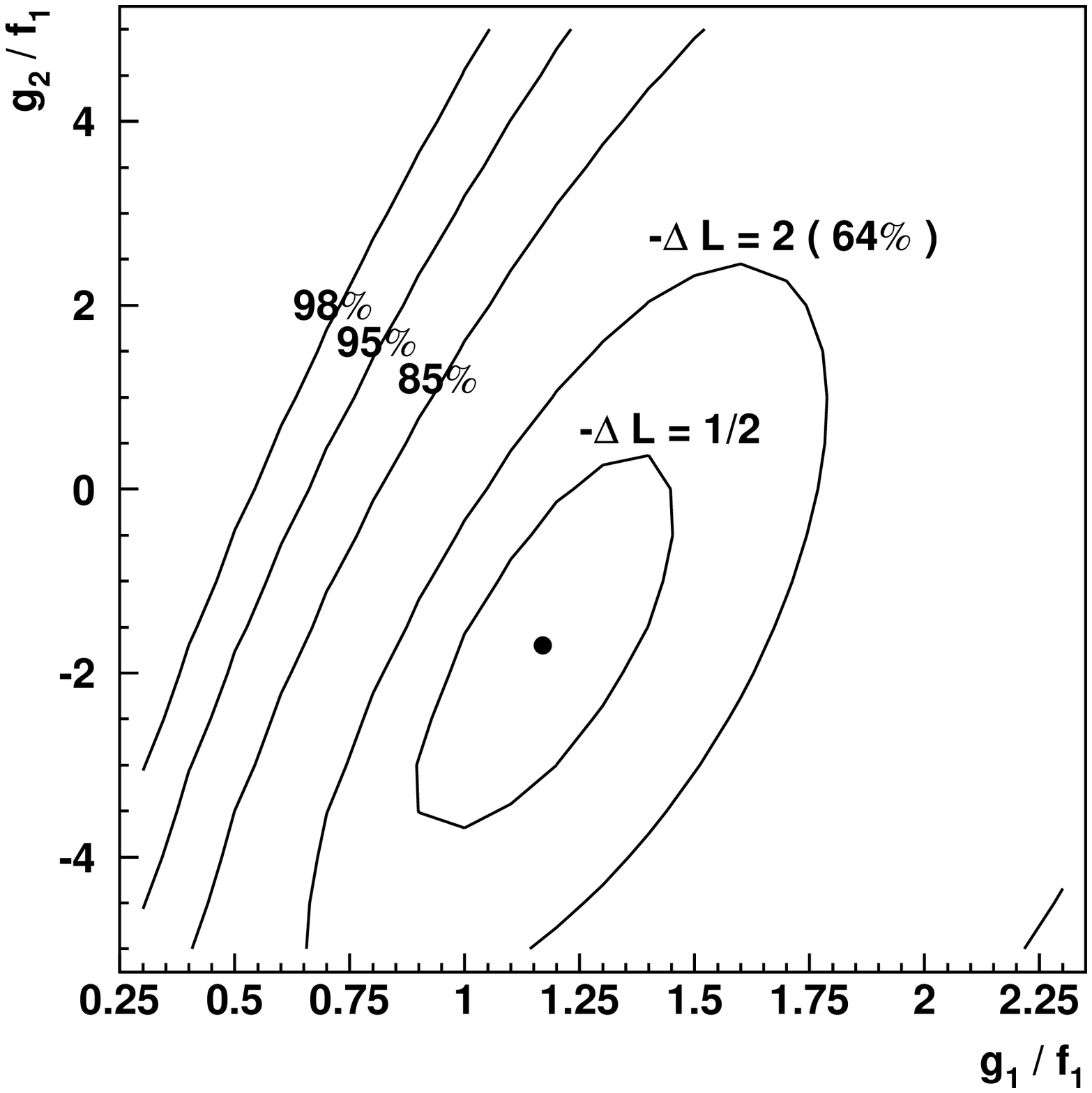}
\caption{ Maximum likelihood fit to \scf and \gf. }
\label{fig:scf}
\end{center}
\end{figure}

Relaxing the requirement that $g_2 = 0$, and fitting the distributions
to \gf and \scf\, simultaneously, one finds no evidence for a
non-zero second class current term (Fig. \ref{fig:scf}),
measuring $ \gf = 1.17 \pm {0.28} ( stat ) \pm 0.05 ( syst ) $
and $ \scf = -1.7 \pm^{2.1}_{2.0} ( stat ) \pm 0.5 ( syst ) $.

Using the measured \gf\!, and assuming $\scf=0$,
one then determines the value for \wf using 
the electron energy spectrum in the \spls 
frame ( Fig. \ref{fig:mk1d}).  (The electron spectrum is the only
kinematic quantity that depends on
\wf to lowest order in $(M_{\Xi^{0}}-M_{\Sigma^{+}})/M_{\Xi^{0}}$.)
A maximum likelihood fit yelds the value \wf = \wfans.

To summarize, the KTeV experiment made the first measurement of \gf for the 
decay \cb\!\!, and found that $\gf = \gfans$ assuming that no
second class current is present and that the weak
magnetism term has the exact \suf value.
By using
the electron--neutrino correlation and the final state polarization
of the \spls observed via its two body decay \sppi,
one was able to determine
both the sign and magnitude of \gf\!.  The answer
is consistent with the exact \suf prediction,
and \suf breaking schemes in which only $g_1$
is allowed to be modified from its exact \suf value.
Predictions that allow for the renormalization of
$f_1$ are disfavored.  Furthermore,
removing the constraint that $g_2 = 0$, and simultaneously fitting
for \gf and \scf\!\!, reveals no evidence for
a second-class current term. The analysis of the energy
spectrum of the electron in the \spls frame
gives a value for \wf that is consistent
with the \suf prediction.

\begin{figure}[tttt]
\epsfxsize=14.0cm
\begin{center}
%s \epsfxsize=20.cm
\epsfbox{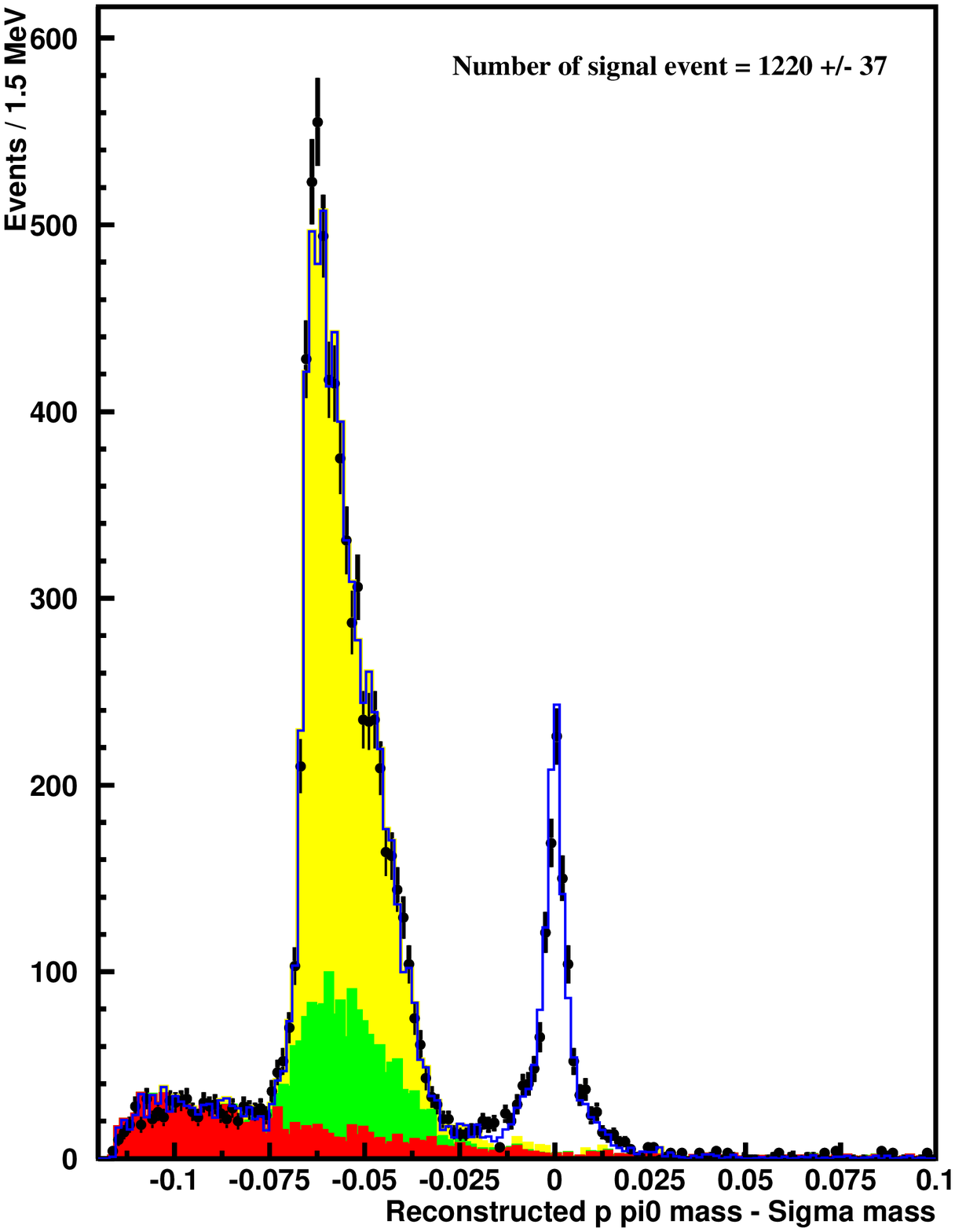}
\caption[The \sppi mass peak]{ The \sppi\, mass peak after all selection 
criteria have been applied.
The background to the left of the peak is due
to \clpi decays ( followed by \lppi or \lb ).
Since \cb is the only source of \spls
in the beam (\csp is kinematically forbidden), 
signal events are identified by
having a $p$-\piz mass within 15 $\mathrm{MeV}$ 
of the nominal \spls mass.}
\label{fig:cas99}
\end{center}
\end{figure}

\begin{figure}[tttt]
\epsfxsize=14.0cm
\begin{center}
%s \epsfxsize=20.cm
\epsfbox{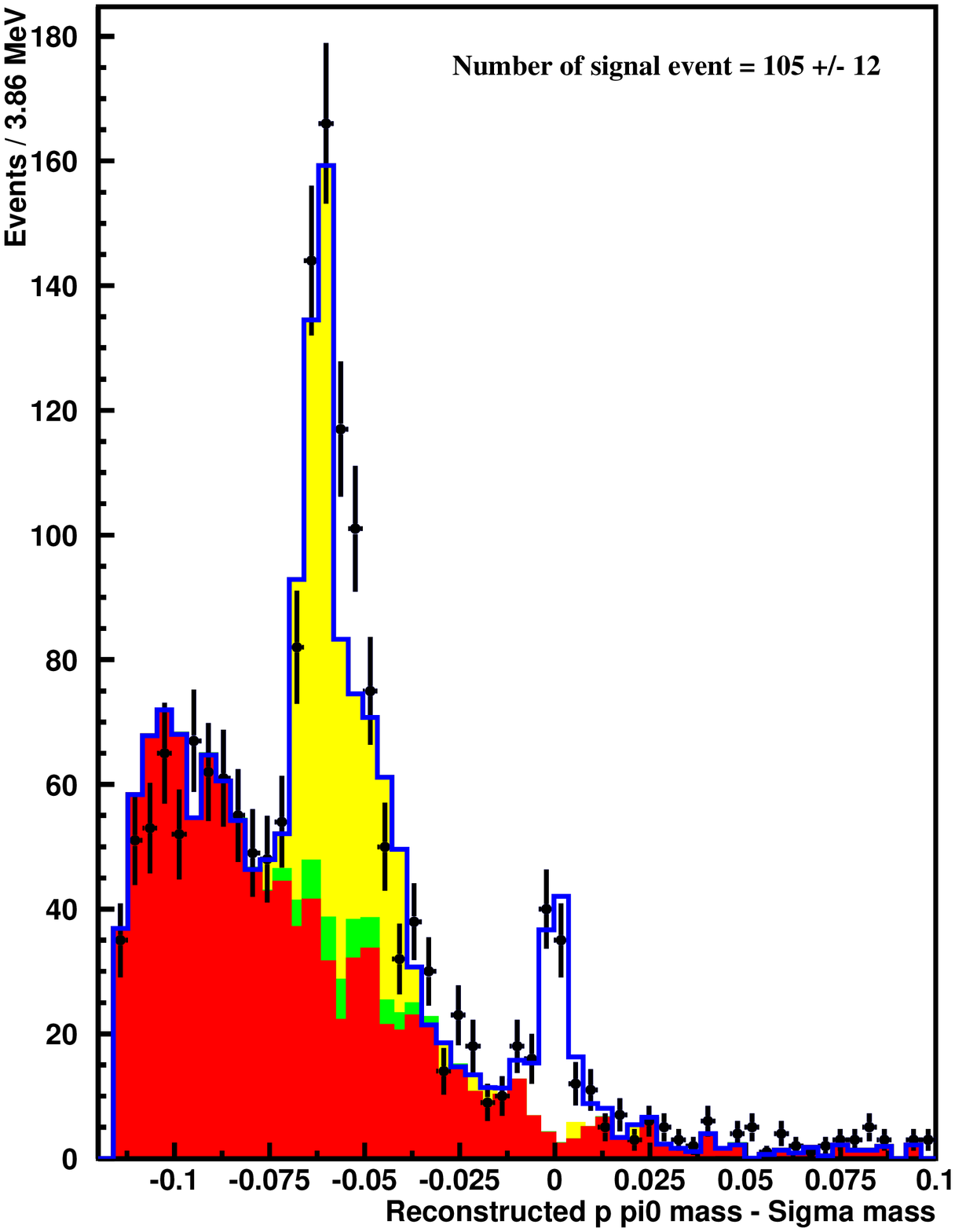}
\caption{$\bar\Sigma^+ $ mass plot from $\bar\Xi^0$ beta decay after all selection 
criteria have been applied.
The background to the left of the peak appears larger as a result of the smaller cross section
for anti \cas production.}

\label{fig:acas99}
\end{center}
\end{figure}

Outlook and future prospects:
(a) \cb:
The KTeV experiment collected a factor of four more data in 1999. 
While analysis is still in progress,  one can already get a sense of the data quality
from the \spls mass plot shown in Fig. \ref{fig:cas99}.
Similarly, one gets an appreciation of the improved sensitivity from a mass plot of 
anti \spls shown in Fig. \ref{fig:acas99}. This is a first observation of anti \cas beta decay. 

The CERN NA 48 experiment reported a sample of 60 \cas beta decay candidates at the 
2000 High Energy Physics Conference in Osaka. Based on this, a succeeding data run 
with somewhat improved instrumentation and trigger is planned for 2002-2004 with the 
expectation of collecting some 25,000 beta decay events. 
This will represent a significant advance in the precision study of hyperon beta decay. 

(b) \lb:
An important advantage of working with intense neutral beams is the possibility of using ``tagged''  $\Lambda$'s from \clpi decays to study
$\Lambda^0$ beta decay, with the double advantage of mitigating the presence of a $Ke_3$ background and of  $\Lambda$'s which are 40\% polarized. The KTeV data from 1999 are expected to have some 5,000 \lb from this source. 
In addition the forthcoming NA48 run at CERN in 2002-2004 can increase this sample by an order of magnitude. Thus  a new level of precision in the \lb 
parameters appears to be within reach.

\subsection{Neutron decay} 

Measurements of neutron beta decay date back to the classic experiments of 
Robson with unpolarized neutrons  \cite{Robson} and of Burgy, Krohn, Novey, Ringo, and Telegdi with polarized neutrons \cite{Burgy}.  Modern experiments 
have measured the neutron lifetime and decay distribution parameters
\cite{PDG2002} with levels of precision much greater than those achieved in hyperon decay experiments. 
First-order recoil effects (arising from terms containing $f_2$ and $g_2$) may even be detectable \cite{Gardner} in the near future.  
At present, six precise measurements of the neutron lifetime give consistent results yielding an average 
of $(885.7 \pm 0.8) $s \cite{PDG2002, Pendelbury}.  Clearly the opportunity to study 
trapped neutrons has greatly advanced the ability to perform these measurements.  

Five contemporary experiments with polarized neutrons yield an average of 
$\alpha_e = -0.1162 \pm 0.0013$ and a form factor ratio $(g_1/f_1)_n = 1.2670 \pm 0.0030$ \cite{PDG2002}.  
These five measurements are not statistically consistent ($\chi^2/d.f. = 10.5/4$) leading to an error scale 
factor of S = 1.6.  The inconsistency arises from the conflict between the published 
PERKEO II result $(g_1/f_1)_n = 1.274 \pm 0.003$ \cite{Abele} and the other four results \cite{Liaud}
(average of $(g_1/f_1)_n = 1.2637 \pm 0.0021$).  A new PERKEO II measurement \cite{AbeleII} exactly confirms
 their earlier result and yields $(g_1/f_1)_n= 1.2739  \pm 0.0019$.  The new experiment has the particular merit 
that the total correction to the raw data is only 2\%, some 10 times smaller than in earlier experiments.  
This new result reinforces the inconsistency.  Combining it with the other five results,
 we obtain an average of $(g_1/f_1)_n = 1.2680 \pm 0.0026$ with $\chi^2/d.f. = 12.4/5 $ and S = 1.6.  
When comparing hyperon beta decay results to the Cabibbo model, the value of $(g_1/f_1)_n$ provides us 
with a powerful constraint : $F + D = (g_1/f_1)_n = 1.2670 \pm 0.0030$. For consistency, we choose to use the 2002 
Particle Data Group average.  In this context the difference between the PERKEO II results and those 
of earlier experiments is of little consequence. On the other hand, this difference is relatively important 
when determining $V_{ud}$ from the neutron lifetime and $(g_1/f_1)_n$ \cite{AbeleII, Towner}.

%end of casbrev.tex

\section{Cabibbo-Model Fits}

The Ademollo-Gatto Theorem suggests an analytic approach to the 
available data that first examines the vector form factor $f_{1}$ 
because it is not subject to first-order  \SU  symmetry breaking 
effects. An elegant way to do this is to use the \textbf{measured} 
value of $g_{1}/f_{1}$ along with the predicted values of $f_{1}$ 
and 
$f_{2}$ (see Table \ref{Table:su3}) to extract a $V_{us}$ value from the decay rate 
for each decay. Consistency of the $V_{us}$ values obtained from 
different decays then indicates the success of the Cabibbo model.

Four hyperon beta decays have sufficient data to perform this 
analysis: $\Lambda \rightarrow p\,e^-\bar\nu,\; 
 \Sigma^- \rightarrow n\,e^-\bar\nu,\; 
\Xi^-\rightarrow \Lambda\,e^-\bar\nu,\;  
\Xi^{0}\rightarrow \Sigma^+\,e^-\bar\nu$. 
Table \ref{Table:Vus} shows the results for them. 
In this analysis, both model-independent and model-dependent radiative 
corrections \cite{GarciaBook} are applied and q$^2$ variation of f$_{1}$ 
and g$_{1}$ is included.  Also \SU values g$_{2}$ = 0 
and f$_{2}$ as given in Table \ref{Table:su3} are used along with the numerical rate 
expressions tabulated in Ref. \cite{GarciaBook}.  The four values are clearly consistent 
($\chi^{2} = 2.26/3$d.f.) with the combined value of $V_{us}$ = 0.2250 
\ensuremath{\pm} 0.0027. This value is nearly as precise as that obtained from 
kaon decay ($V_{us}$ = 0.2196 \ensuremath{\pm} 0.0023) and, as 
suggested by previous analyses \cite{Gaillard:1984ny, highvus, 
Flores-Mendieta:1998ii}, is somewhat larger. In combination 
with $V_{ud}$ = 0.9740 \ensuremath{\pm} 0.0005 obtained from 
superallowed pure Fermi nuclear decays \cite{Towner}, the larger $V_{us}$ 
value from hyperon decays beautifully satisfies the unitarity constraint 
{\textbar}$V_{ud}${\textbar}$^{2}$ + 
{\textbar}$V_{us}${\textbar}$^{2}$ + 
{\textbar}$V_{ub}${\textbar}$^{2}$ 
= 1. As discussed extensively by Towner and Hardy \cite{Towner}, the 
unitarity constraint falls short by 2.2 standard deviations with 
the smaller kaon decay value. 

% This is the Vus Analysis Table
\begin{table}[htb]
\begin{center}
\caption{Results from $V_{us}$ analysis using measured $g_1/f_1$ values}
\begin{tabular}{@{}lccc@{}}
\hline
\hline
Decay          &        Rate        &   $g_1/f_1$   &   $V_{us}$   \\  
Process        &  ($\mu sec^{-1}$)  &               &       \\  
\hline
$\Lambda \rightarrow p e^- \overline{\nu}$ &  $3.161(58)$      &  $0.718(15)$  &  0.2224 $\pm$ 0.0034  \\
$\Sigma^- \rightarrow n e^- \overline{\nu}$ & $6.88(24)$       &  $-0.340(17)$ &  0.2282 $\pm$ 0.0049   \\
$\Xi^- \rightarrow \Lambda e^- \overline{\nu}$ & $3.44(19)$    &  $0.25(5)$     &  0.2367 $\pm$ 0.0099  \\
$\Xi^0 \rightarrow \Sigma^+ e^- \overline{\nu}$ & $0.876(71)$  &  $1.32(+.22/-.18)$  &  0.209 $\pm$ 0.027  \\
Combined   &    ---    &     ---     &     0.2250 $\pm$ 0.0027 \\
\hline
\end{tabular}
\label{Table:Vus}
\end{center}
\end{table}

First order  \SU  symmetry breaking effects are expected to manifest 
themselves in g$_{1}$/f$_{1}$. The newly measured decay 
$\Xi^{0}\rightarrow \Sigma^+\,e^-\bar\nu$ provides a direct test 
because it is predicted 
to have the same form factor ratio as the well-measured neutron 
beta decay: $n \rightarrow p\,e^-\bar\nu$. As shown in Table \ref{Table:datasum}, 
the results 
are consistent with this prediction, but the errors are currently 
rather large. For the other decays, it is necessary to fit for 
the reduced axial-vector form factors $F$ and $D$. Since $g_{1}/f_{1} 
= F + D$ for neutron beta decay, this combination is very well 
determined. It is therefore better to fit for the linear combinations 
$F + D$ and $F - D$ which will then have essentially uncorrelated 
errors. This fit yields $F + D = 1.2670 \,\pm 0.0030$ and 
$F - D = -0.341 \pm 0.016$ with $\chi^{2} = 
2.96/3$ d.f. As 
might be expected, the result for F - D is dominated by the 
reasonably 
precise $g_{1}/f_{1}$ value for 
$ \Sigma^- \rightarrow n\,e^-\bar\nu$. 
Surprisingly, even with today's improved measurements, no clear 
evidence of  \SU  symmetry breaking effects emerges. They appear 
to be much smaller than expected.

The value for $V_{us}$ determined from hyperon beta decays, 
without applying any \SU breaking corrections to $f_1$, but including 
radiative corrections, is:
\begin{equation}
\label{VusHyp}
V_{us} = 0.2250 \,(27) \quad \text{hyperon beta decay}
\end{equation}
The expected negative correction to $f_1$ would drive this higher, to
\begin{align}
\label{VusHypCorr}
  V_{us}    &   = 0.2279\, (27)\quad \text{Ref. \cite{Donoghue:th}} \\
     V_{us} &  =  0.2307\, (27)\quad \text{Ref. \cite{Schlumpf:1994fb}}\nonumber
\end{align}
The accepted value from $K_{l 3}$ decays,  including 
corrections from chiral perturbation theory, \cite{Leutwyler},  is
considerably lower,
\begin{equation}
\label{VusK}
V_{us} = 0.2196 \,(23) \quad K_{l 3}\text{ decay}
\end{equation}
We have a puzzle: why are the two values different? If we assume
that the $f_1$ correction in hyperon beta decays must be negative,
Eq. \eqref{VusHyp} must be considered a lower limit for $V_{us}$. 
We are thus driven to the conclusion that the value from  $K_{l 3}$ is 
an underestimate. Is this possible? Perhaps yes: a quark-model
computation \cite{Jaus} of the $F_+(0)$ correction in $K_{l 3}$ 
finds $F_+(0)=0.965$. While at first sight this is compatible with
the chiral perturbation theory  \cite{Leutwyler} result, $F_+(0)=0.961$,
 it is not evident that
the two computations reflect \emph{the same corrections}. It could well be
that the two corrections should, at least in part, be combined. 
In other words it is possible that chiral perturbation theory
neglects some short-distance contribution 
which is well described by the quark model, and this would 
explain the discrepancy between the two results.

While it is clear that more theoretical work is needed to fully clarify the situation,
we are at this point convinced that there is no reason to prefer the $K_{l 3}$
result over the one derived from hyperon beta decays.  Indeed there is now also a 
preliminary experimental indication \cite{Sher} that the $K_{e 3}$ decay rate may be 
higher than the value used to obtain Eq. \eqref{VusK}.  

\section{Conclusions and Open Questions}

The determination of  the elements of the CKM matrix is one of the main
ingredients for evaluating the solidity of the standard model of elementary particles.
This is a vast subject which has seen important progress with the  determination of 
$\epsilon'/\epsilon$ and the observation of  $CP$ violation in B decays.

While a lot of attention has recently  been justly devoted to the higher mass sector of
the CKM matrix, it is the low mass sector, in particular $V_{ud}$ and $V_{us}$ where 
the highest precision  can be attained, and which can provide the most sensitive test 
of the unitarity of the CKM matrix through the relation 
$|V_{ud}|^2+|V_{us}|^2+|V_{ub}|^2=1$.  Given the fact that the $|V_{ub}|^2$
contribution is totally negligible,  the unitarity test reduces to the consistency of 
$\cos \theta_C$ determined from nuclear beta decay and  of  $\sin \theta_C$ 
determined from strangeness changing semileptonic decays.

In the present review we have reconsidered the contribution that  the hyperon beta 
decays can give to the determination of   $\sin \theta_C$. The conventional analysis
of hyperon beta decay  in terms of the $F, D$ and $\sin \theta_C$ parameters is marred
by the expectation of first order $SU(3)$ breaking effects in the axial-vector
contribution. The situation is only made worse if one introduces adjustable 
$SU(3)$ breaking parameters as this  increases the number of degrees of freedom and
degrades the precision. If on the contrary, as we did here, one focuses the analysis 
on the vector form factors, treating the rates and $g_1/f_1$ as the basic experimental data, 
one has direct access to the $f_1$  form factor  for each decay, and this in turn allows 
for a redundant determination of  $\sin \Tc$. The consistency of the values of  $\sin \Tc$ 
determined from the different decays is a first confirmation of the overall consistency
of the model.

The value of   $\sin \Tc$ obtained from hyperon decays is of comparable 
precision with that obtained from $K_{l3}$ decays, and is in better agreement
with the value of \Tc obtained from nuclear beta decay. While a discrepancy 
between $V_{us}$ and $V_{ud}$ could be seen as a portent of exciting new physics, 
a discrepancy between the two different determinations of   $V_{us}$ can only be 
taken as an indication that more work remains to be done both on the theoretical and
the experimental side. 

On the theoretical side, renewed efforts are needed for the determination 
of  \SU\!\!-breaking effects in hyperon beta decays as well as in $K_{l 3}$ 
decays. While it is 
quite possible to improve the present situation on the quark-model 
front, the best hopes lie in  lattice 
QCD simulations, perhaps combined with chiral perturbation theory
for the evaluation of large-distance multiquark contributions.

We have given some indication that  the trouble could arise from the 
$K_{l 3}$ determination of $\sin\Tc$, and we would like to encourage further experimental 
work in this field. We are however convinced of  the importance of
renewed experimental work on hyperon decays, of the kind now in progress
at the CERN SPS. The interest of this work goes beyond  the determination of
$\sin\Tc$, as it involves  the intricate and elegant relationships that the
model predicts.

\section{Acknowledgements}
With permission from the Annual Review of Nuclear and Particle Science. 
Final version of this material is scheduled to appear in the Annual Review of Nuclear and Particle Science Vol. 53, 
to be published in December 2003 by Annual Reviews, http//AnnualReviews.org.
The idea for undertaking this review arose during Hyperon 99 at Fermilab. We are grateful to 
D. A. Jensen and E. Monnier for organizing this stimulating conference. 
The continuing intellectual stimulation provided by colleagues in the Fermilab KTeV Collaboration, particularly members 
of the hyperon working group, is gratefully acknowledged.  We thank J. L. Rosner for a thorough and thoughtful 
reading of our manuscript.  This work was supported in part by the U.S. Department of Energy under grant 
DE-FG02-90ER40560 (Task B).

\appendix
\section{Appendix}
We present the analytic expressions for the integrated
final state polarization
to order $ \delta^2 $ in the final state rest frame, 
assuming real form factors.  

\begin{eqnarray}
R & = &  R_{0} [ ( 1 - \frac{3}{2} \delta ) \fisq
      +  ( 3 - \frac{9}{2} \delta ) \gisq 
      -  ( 4 \delta ) \gigj ] + R(\delta^2) , \nonumber \\
R \mathsf{S}_{e}  & = &  R_{0} [ ( 2 - \frac{10}{3} \delta ) \gisq
      +  ( 2 - \frac{7}{3} \delta ) \figi
      -  (\frac{1}{3} \delta ) \fisq \nonumber  \\
& &
      -  (\frac{2}{3} \delta ) \fifj 
      +  (\frac{2}{3} \delta ) \fjgi
      -  (\frac{2}{3} \delta ) \figj
      -  (\frac{10}{3} \delta  ) \gigj ]
      + R \mathsf{S}_{e} (\delta^2) , \nonumber \\
R  \mathsf{S}_{\nu} & = &  R_{0} [( -2 + \frac{10}{3} \delta ) \gisq
      +  ( 2 - \frac{7}{3} \delta ) \figi
      +  ( \frac{1}{3} \delta ) \fisq \nonumber  \\
& &
      +  ( \frac{2}{3} \delta ) \fifj 
      +  ( \frac{2}{3} \delta ) \fjgi
      -  ( \frac{2}{3} \delta ) \figj
      +  ( \frac{10}{3} \delta ) \gigj ]
      + R \mathsf{S}_{\nu} (\delta^2) , \nonumber \\
R \mathsf{S}_{\alpha}  & = & R_{0}
[( \frac{8}{3} - \frac{52}{15} \delta ) \figi
      +  (\frac{16}{15} \delta ) \fjgi
      -  (\frac{16}{15} \delta ) \figj ]
      + R \mathsf{S}_{\alpha} (\delta^2), \nonumber \\
R \mathsf{S}_{\beta}  & = &  R_{0} [( \frac{8}{3} - 4 \delta )\gisq
      -  (\frac{8}{15} \delta ) \fisq
      -  (\frac{16}{15} \delta ) \fifj  \nonumber \\
& &
      -  (\frac{64}{15} \delta ) \gigj ]
      + R \mathsf{S}_{\beta} (\delta^2), 
\end{eqnarray}

where 

\begin{eqnarray}
R_{0} = \frac{G_{S}^{2} (\delta M_{B})^{5}}{60 \pi^{3}}. \nonumber
\end{eqnarray}

\begin{eqnarray}
R(\delta^2) & = &   R_{0} \delta^2 (\frac{6}{7} \fisq
+ \frac{12}{7} \gisq + 6 \gigj  \nonumber \\
& &
+ \frac{6}{7} \fifj + \frac{4}{7} \fjsq + \frac{12}{7} \gjsq ),\nonumber \\
R \mathsf{S}_{e}(\delta^2) & = &  R_{0} \delta^2 ( \frac{55}{42} \gisq
+ \frac{17}{21} \figi + \frac{19}{42} \fisq
+ \frac{4}{3} \fifj - \frac{10}{21} \fjgi \nonumber \\
& & 
+ \frac{10}{21} \figj + \frac{116}{21} \gigj 
+ \frac{4}{21} \fjsq + \frac{4}{3} \gjsq -  \frac{16}{21} \fjgj ), \nonumber \\
R \mathsf{S}_{\nu}(\delta^2) & = &  R_{0} \delta^2 ( -\frac{55}{42} \gisq
+  \frac{17}{21} \figi - \frac{19}{42} \fisq
- \frac{4}{3} \fifj - \frac{10}{21} \fjgi \nonumber \\
& &
+ \frac{10}{21} \figj - \frac{116}{21} \gigj 
-  \frac{4}{21} \fjsq - \frac{4}{3} \gjsq -  \frac{16}{21} \fjgj ), \nonumber \\
R \mathsf{S}_{\alpha}(\delta^2) & = &  R_{0} \delta^2 ( \frac{316}{245} \figi
- \frac{752}{735} \fjgi + \frac{752}{735} \figj 
- \frac{128}{105} \fjgj  ), \nonumber \\
R \mathsf{S}_{\beta}(\delta^2) & = &  R_{0} \delta^2 ( 
\frac{422}{735} \fisq + \frac{88}{49} \fifj + \frac{8}{35} \fjsq  \nonumber \\
& &
+ \frac{362}{245} \gisq + \frac{1576}{245} \gigj + \frac{8}{5} \gjsq )
\end{eqnarray}
%\listoffigures
%\listoftables


\begin{thebibliography}{99}

%\cite{cab}
\bibitem{cab}
Cabibbo N. \textit{Phys. Rev. Lett.} 10:531 (1963)

\bibitem{km} 
Kobayashi M, Maskawa T. \textit{Prog. Theor. Phys.} 49:652 (1973)

\bibitem{cb} Affolder A, et al (KTeV Collaboration). \textit{Phys. Rev. Lett.} 82:3751 (1999)

\bibitem{ffcb} 
Alavi-Harati A, et al (KTeV Collaboration). \textit{Phys. Rev. Lett.} 87:132001 (2001)

%\cite{OkunRochester58}
\bibitem{OkunRochester58}
Okun L. In \textit{Proc. 1958 Int. Conf. on High Energy Physics at CERN, June 30--July 5, 1958},  
ed. B. Ferretti, p. 223. Geneva: CERN (1958)

%\cite{SakataModel}
\bibitem{SakataModel}
%SakataModel-To be found%%% Da fare!
Sakata S. \textit{Prog. Theor. Phys.} 16:686 (1956); see also, 
Maki Z, Nakagawa M, Sakata S. \textit{Prog. Theor. Phys.} 28:870 (1962)

%\cite{Berman}
\bibitem{Berman}
Berman SM. \textit{Radiative Corrections to Muon and Neutron 
Decay}. PhD Thesis. California Institute of Technology, Pasadena, 
California, (1959);
Berman SM. \textit{Phys. Rev.} 112:267 (1958);
Kinoshita T, Sirlin A.
\textit{Phys. Rev.} 113:1652 (1959)

%\cite{Kinoshita:1958ru}
%\bibitem{Kinoshita:1958ru}
%``Radiative Corrections To Fermi Interactions,''
%%CITATION = PHRVA,113,1652;%%


\bibitem{Gell-Mann:np}
Gell-Mann M, L\'{e}vy M.
%``The axial-vector Current In Beta Decay,''
\textit{Nuovo Cimento} 16:705 (1960)
%%CITATION = NUCIA,16,705;%%


%\cite{AngelaBG}
\bibitem{AngelaBG}
Barbaro-Galtieri A, et al. 
\textit{Phys. Rev. Lett.} 9:26 (1962)


%\cite{Cronin68}
\bibitem{Cronin68}
Cronin J. In \textit{Proc. 14th Int. Conf. on High Energy Physics, 
Vienna, Aug. 28--Sept. 5, 1968}, ed. Prentki J, Steinberger J, p. 281. 
Geneva: CERN (1968) 
%1968 "Rochester Conference"

%\cite{PDG2002}
\bibitem{PDG2002}
Hagiwara K, et al (Particle Data Group). \textit{Phys. Rev.} D66:010001 (2002)

%\cite{GellMannCA}
\bibitem{GellMannCA}
Gell-Mann M. 
\textit{Phys. Rev.} 125:1067 (1962)

%\cite{AdlWeisbrg}
\bibitem{AdlWeisbrg}
Weisberger WI. \textit{Phys. Rev. Lett.} 14:1047 (1965);
Adler SL. \textit{Phys. Rev. Lett.} 14:1051 (1965)

%\cite{gla_cab}
\bibitem{gla_cab}
Sheldon Glashow in a private communication with
N Cabibbo

%\cite{KTeVepsi}
\bibitem{KTeVepsi}
Alavi-Harati A, et al (KTeV Collaboration).
\textit{Phys. Rev.} D66:11????? (2002)
[arXiv:hep-ex/0208007]  {???help???}
%KTeVepsi -To be done %%% da completare


%\cite{Lai:2001ki}
\bibitem{Lai:2001ki}
Lai A, et al (NA48 Collaboration).
%``A precise measurement of the direct CP violation parameter  \textrm{Re}( epsilon'/epsilon),''
\textit{Eur. Phys. J.} C22:231 (2001)
[arXiv:hep-ex/0110019]
%%CITATION = HEP-EX 0110019;%%

%\cite{Babar}
\bibitem{Babar}
Aubert B, et al (BaBar Collaboration). Contributed to \textit{31st Int. Conf. on High 
Energy Physics (ICHEP 2002), Amsterdam, The Netherlands, July 24--31, 2002} [arXiv:hep-ex/0207070]
%Babar -To be done %%% da completare

%\cite{Belle}
\bibitem{Belle}
Abe K, et al (Belle Collaboration). Contributed to \textit{31st Int. Conf. High 
Energy Physics (ICHEP 2002), Amsterdam, The Netherlands, July 24--31, 2002} [arXiv:hep-ex/0207098]

%Belle -To be done %%% da completare


%\cite{CabibboLP01}
\bibitem{CabibboLP01}
Nir Y. Plenary talk \textit{31st Int. Conf. High 
Energy Physics (ICHEP 2002), Amsterdam, The Netherlands, July 24--31, 2002} [arXiv:hep-ph/0208080]
%N. Cabbibo 
%To be done  %%% da completare


%\cite{Gaillard:1984ny}
\bibitem{Gaillard:1984ny}
Gaillard JM, Sauvage G.
%``Hyperon Beta Decays,''
\textit{Annu. Rev. Nucl. Part. Sci.} 34:351 (1984)
%%CITATION = ARNUA,34,351;%%

\bibitem{bri} Bright S, Winston R, Swallow EC, Alavi-Harati A.
\textit{Phys. Rev.} D60:117505 (1999); 
see also Watson JM and Winston R. \textit{Phys. Rev.} 
181:1907 (1969) 


% \cite{GarciaBook}
\bibitem{GarciaBook}
Garcia A, Kielanowski R. \textit{The Beta Decay of Hyperons}, 
Lecture Notes in Physics 222. Berlin: Springer-Verlag (1985)


% \cite{BjDrell}
\bibitem{BjDrell}
Bjorken JD, Drell SD. 
\textit{Relativistic Quantum Mechanics}. 
New York: McGraw-Hill (1964)


%\cite{WeinbergII}
\bibitem{WeinbergII}
Weinberg S. 
\textit{Phys. Rev.} 112:1375 (1958)


\bibitem{pri1} Primakoff H. \textit{Rev. Mod. Phys.} 31:802 (1959)

\bibitem{pri2} Fujii A, Primakoff H. \textit{Nuovo Cimento} 12:327 (1959)

%\cite{Weinbergthm}  
\bibitem{Weinbergthm}
Weinberg S. 
\textit{Phys. Rev.} 115:481 (1959)

%\cite{Sirlin}
\bibitem{Sirlin}
Sirlin A. \textit{Phys. Rev.} 164:1767 (1967); 
\textit{Phys. Rev. Lett.} 32:966 (1974)

%\cite{radcor}
\bibitem{radcor}
Most recently, Martinez A, Torres JJ, Flores-Mendieta R, Garcia A.
\textit{Phys. Rev.} D63:014025 (2001) and references therein


%\cite{Sirlinf2}
\bibitem{Sirlinf2}
Sirlin A. \textit{Nucl. Phys.} B161:301 (1979) 

%\cite{ad}
\bibitem{ad} Ademollo M, Gatto R. 
\textit{Phys. Rev. Lett.} 13:264 (1964)

%\cite{fufu} 
\bibitem{fufu}
Fubini S, Furlan G. \textit{Physics} 1:229 (1965)

%\cite{SakuraiAG} 
\bibitem{SakuraiAG}
Sakurai J. J. 
\textit{Currents 
and Mesons}, Chicago Lectures in Physics. Chicago:University of Chicago 
Press (1969)

%\cite{Donoghue:th}
\bibitem{Donoghue:th} Donoghue JF, Holstein BR, Klimt SW. \textit{Phys. Rev.} D35:934 (1987)
%``K-M Angles And SU(3) Breaking In Hyperon Beta Decay,''
%%CITATION = PHRVA,D35,934;%%


%\cite{Schlumpf:1994fb}
\bibitem{Schlumpf:1994fb}
Schlumpf F.
%``Beta decay of hyperons in a relativistic quark model,''
\textit{Phys. Rev.} D51:2262 (1995)
[arXiv:hep-ph/9409272]
%%CITATION = HEP-PH 9409272;%%

%\cite{Krause:xc}
\bibitem{Krause:xc}
Krause A.
%``Baryon Matrix Elements Of The Vector Current In Chiral Perturbation Theory,''
\textit{Helv. Phys. Acta} 63:3 (1990)
%%CITATION = HPACA,63,3;%%

%\cite{Anderson:1993as}
\bibitem{Anderson:1993as}
Anderson J, Luty MA.
%``Chiral corrections to hyperon vector form-factors,''
\textit{Phys. Rev.} D47:4975 (1993)
[arXiv:hep-ph/9301219]
%%CITATION = HEP-PH 9301219;%%


%\cite{quenched}
\bibitem{quenched}
Sexton J, Weingarten D.
%"Error estimate for the valence approximation and for a systematic expansion of full QCD"
\textit{Phys. Rev.} D55:4025 (1997)
%[arXiv:hep-ph/9301219].

%To be found %%%


%\cite{Flores-Mendieta:1998ii}
\bibitem{Flores-Mendieta:1998ii}
Flores-Mendieta R, Jenkins E, Manohar AV.
%``SU(3) symmetry breaking in hyperon semileptonic decays,''
\textit{Phys. Rev.} D58:094028 (1998)
[arXiv:hep-ph/9805416]
%%CITATION = HEP-PH 9805416;%%


%\cite{Quinn:1968qy}
\bibitem{Quinn:1968qy}
Quinn HR, Bjorken JD.
%``Renormalization Of Weak Form-Factors, And The Cabibbo Angle,''
\textit{Phys. Rev.} 171:1660 (1968)
%%CITATION = PHRVA,171,1660;%%

%\cite{Holstein}
\bibitem{Holstein}
Holstein BR. In \textit{Hyperon 99, Proc. Hyperon Phys. Symp., Fermilab, Sept. 27-29 
1999}, ed. DA Jensen, E Monnier, p 4. Fermilab report 
FERMILAB-Conf-00/059-E (2000) 

%\cite{E715}
\bibitem{E715}
Hsueh SY, et al. \textit{Phys. Rev.} D38:2056 (1988)


%\cite{Oehme}
\bibitem{Oehme}
Oehme R, Winston R, Garcia A. \textit{Phys. Rev.} D3:1618 (1971)

%\cite{HypBeams}
%\bibitem{HypBeams}
%To be done%%%

%\cite{CabChil}
\bibitem{CabChil}
Cabibbo N, Chilton F. \textit{Phys. Rev.} B137:1628 (1965)

%\cite{Bourquin:1983III}
\bibitem{Bourquin:1983III}
Bourquin M, et al. \textit{Z. Phys.} C21:17 (1983); 
see also Tanenbaum W, et al. \textit{Phys. Rev.} D12:1871 (1975) 

%\cite{Keller}
\bibitem{Keller}
Keller P, et al. \textit{Phys. Rev. Lett.} 48:971 (1982) and references therein;
see also Garcia A, Swallow EC. \textit{Phys. Rev. Lett.} 35:467 (1975)

%\cite{GarciaBohm}
\bibitem{GarciaBohm}	Discussed at length in  \cite{GarciaBook};  
see, for example, 
Bohm A, Magnollay P, Garcia A, Kielanowski P. \textit{Phys. Rev.} D27:180 (1983); 
Bohm A, Kmiecik M. \textit{Phys. Rev.} D31:3005 (1985)

%\cite{Bunce}
\bibitem{Bunce}
First observed for neutral hyperons: 
Bunce G, et al. \textit{Phys. Rev. Lett.} 36:1113 (1976)

%\cite{Pondrom}
\bibitem{Pondrom}
Lach J, Pondrom LG. \textit{Annu. Rev. Nucl. Part. Sci.} 29:203 (1979)

%\cite{Zapalac} 
\bibitem{Zapalac}
Zapalac G, et al. \textit{Phys. Rev. Lett.} 57:1526 (1986)

%\cite{WiseLambda}
\bibitem{WiseLambda}
Wise J, et al. \textit{Phys. Lett.} B91:165 (1980); 
Wise J, et al. \textit{Phys. Lett.} B98:123 (1981) 

%\cite{PDG78}
\bibitem{PDG78}
Bricman C, et al (Particle Data Group). \textit{Phys. Lett.} B75:1 (1978)

% \cite{JensenPrivate}
\bibitem{JensenPrivate}
Jensen DA. private communication

%\cite{Lindquist}
\bibitem{Lindquist}
See, for example, Lindquist J, et al. \textit{Phys. Rev.} D16:2104 (1977)

%\cite{FermiLambda}
%\bibitem{FermiLambda}

%To be found %%%

%\cite{Bourquin:1983I}
\bibitem{Bourquin:1983I}
Bourquin M, et al. \textit{Z. Phys.} C21:1 (1983)

\bibitem{e'ktev}
Alavi-Harati A, et al (KTeV Collaboration). \textit{Phys. Rev. Lett.} 83:22 (1999)

%\bibitem{aahthesis}
%Alavi-Harati A. \textit{Observation and Branching Fraction Measurement of 
%$\Xi^{0}\rightarrow \Sigma^{+}\, e^{-}\,\bar\nu$ at KTeV/E799-II, Fermilab} 
%PhD Thesis. The University of Wisconsin -- Madison (1999)

\bibitem{aahDPF99}
Alavi-Harati A, et al (KTeV Collaboration). In \textit{DPF99: Proc. 1999 
APS/DPF Meeting, Los Angeles, Jan. 5--9, 1999}, ed. K Arisaka, Z Bern, 
http://www.dpf99.library.ucla.edu/session4/alavi0408.pdf 

%To be found %%%

%\bibitem{e'na48}
%To be found %%%

\bibitem{ns} The feasibility was first studied by N. Solomey working with 
 University of Chicago undergraduate A. Affolder


\bibitem{dwo} Dworkin J, et al. \textit{Phys. Rev.} D41:780 (1990)

%neutron
\bibitem{Robson}	Robson J. \textit{Can. J. Phys.} 36:1450 (1958)

\bibitem{Burgy}	Burgy M, Krohn V, Novey T, Ringo G, Telegdi V. \textit{Phys. Rev.} 120:1827 (1960)

%\bibitem{Groom}	D. E. Groom {\it et al.}, (Particle Data Group), Eur. Phys. J. C 15, 1 (2000) and 2001 off-year update at http://pdg.lbl.gov.

\bibitem{Gardner}	Gardner S, Zhang C. \textit{Phys. Rev. Lett.} 86:5666 (2001)

\bibitem{Pendelbury}	For reviews and commentary see 
Pendelbury JM. \textit{Annu. Rev. Nucl. Part. Sci.} 43:687 (1993); 
Schreckenbach J, et al. \textit{J. Phys.} G18:1 (1992); 
Freedman SJ. \textit{Comm. Nucl. Part. Phys.} 19:209 (1990)

\bibitem{Abele}	Abele H, et al. \textit{Phys. Lett.} B407:212 (1997)

\bibitem{Liaud}	Mostovoi YuA, et al. \textit{Phys. Atom. Nucl.} 
64:1955 (2001) [tr. YAF 64:2040 (2001)], 
Liaud P, et al. \textit{Nucl. Phys.} A612:53 (1997); 
Yerozolimsky B, et al. \textit{Phys. Lett.} B412:240 (1997);
Bopp P, et al. \textit{Phys. Rev. Lett.} 56:919 (1986)

\bibitem{AbeleII}	Abele H, et al. \textit{Phys. Rev. Lett.} 88:211801 (2002)

\bibitem{Towner}	Towner IS, Hardy JC. In \textit{Proc. Vth Int. WEIN Symp.: Phys. Beyond the Standard
Model, Santa Fe, NM, 1998}, ed. P Herczeg, C Hoffman, HV Klapdor-Kleingrothaus, p. 338. 
Singapore: World Scientific (1999) [arXiv:nucl-th/9809087]
%\cite{Ratcliffe}
%\cite{rat}
\bibitem{rat} Ratcliffe PG. \textit{Phys. Rev.} D59:014038 (1999)

%\cite{fm}
%\bibitem{fm} Flores-Mendieta R, Jenkins E, Manohar AV. \textit{Phys. Rev.} D58:094028 (1998)

\bibitem{highvus} Flores-Mendieta R, Garcia A, S'anchez-Col'on G. \textit{Phys. Rev.} D54:6855 (1996); 
Dai J, Dashen R, Jenkins E, Manohar AV. \textit{Phys. Rev.} D53:273 (1996)


%\cite{Leutwyler}
\bibitem{Leutwyler}
Leutwyler H, Roos M. \textit{Z. Phys.} C25:91 (1984)
%See also Lopez

%\cite{Sher}
\bibitem{Sher}
Sher A, et al (E865 Collaboration).  Contributed to \textit{DPF2002: 2002 
APS/DPF Meeting, Williamsburg, Virginia, May 24--28, 2002}, 
http://dpf2002.velopers.net/talks\_pdf/332talk.pdf  


%\cite{Jaus} 
\bibitem{Jaus}
Jaus W.  \textit{Phys. Rev.} D44:2851 (1991)
 %+%

%\cite{
%\cite{Okun:vy}
%\bibitem{Okun:vy}
%L.~Okun and B.~Pontecorvo,
%``Some Remarks On Slow Processes Of Transformation Of Elementary  Particles,''
%Zh.\ Eksp.\ Teor.\ Fiz.\  {\bf 32}, 1587 (1957).
%CITATION = ZETFA,32,1587;%%
%
%\cite{SaitoThomas}
%\bibitem{SaitoThomas}
%K. P. Saito and A. W. Thomas, Phys. Lett B363 {\bf 157} (1995)

%\cite{Bourquin:1983fh}
%\bibitem{Bourquin:1983fh}
%M.~Bourquin {\it et al.}  [Bristol-Geneva-Heidelberg-Orsay-Rutherford-
%                  Strasbourg Collaboration],
%Z.\ Phys.\ C {\bf 21}, 27 (1983).
%``Measurements Of Hyperon Semileptonic Decays At The Cern Super Proton Synchrotron. 4. Tests Of The Cabibbo Model,''
%%CITATION = ZEPYA,C21,27;%%
%

%\cite{Martinez:2000wp}
%\bibitem{Martinez:2000wp}
%A.~Martinez, J.~J.~Torres, R.~Flores-Mendieta and A.~Garcia,
%``Radiative corrections to the semileptonic Dalitz plot with angular  correlation between polarized decaying hyperons and emitted charged
%leptons,''
%Phys.\ Rev.\ D {\bf 63}, 014025 (2001)
%[arXiv:hep-ph/0006279].
%%CITATION = HEP-PH 0006279;%%

%
%\cite{Flores-Mendieta:2001nc}
%\bibitem{Flores-Mendieta:2001nc}
%R.~Flores-Mendieta, A.~Garcia, A.~Martinez and J.~J.~Torres,
%``Radiative corrections to the semileptonic Dalitz plot with angular  correlation between polarized decaying and emitted hyperons: Effects of
%the four-body region,''
%Phys.\ Rev.\ D {\bf 65}, 074002 (2002)
%[arXiv:hep-ph/0111117].
%%CITATION = HEP-PH 0111117;%%

%
%\cite{Flores-Mendieta:1996gz}
%\bibitem{Flores-Mendieta:1996gz}
%R.~Flores-Mendieta, A.~Garcia and G.~Sanchez-Colon,
%``Determination of the Kobayashi-Maskawa-Cabibbo matrix element $V_{us}$ under various flavor-symmetry breaking models in hyperon semileptonic
%decays,''
%Phys.\ Rev.\ D {\bf 54}, 6855 (1996)
%[arXiv:hep-ph/9603256].
%%CITATION = HEP-PH 9603256;%%

%

%\cite{Luty:1993jh}


%\bibitem{Luty:1993jh}
%M.~A.~Luty and M.~J.~White,
%``SU(3) versus SU(3) x SU(3) breaking in weak hyperon decays,''
%$arXiv:hep-ph/9304291.
%%CITATION = HEP-PH 9304291;%%
%\end{thebibliography}
% +++ extra references!
%\cite{Ratcliffe}
%\bibitem{rat} P. G. Ratcliffe, Phys. Rev. D {\bf 59} 014038 (1999).
%\cite{F-M}
%\bibitem{fm} R. Flores-Mendieta {\it et al.}, Phys. Rev. D {\bf 58} 094028 (1998).
%
%\bibitem{cb} A.\ Affolder {\it et al.}, Phys. Rev. Lett. {\bf 82}, 3751 (1999).
%
%\bibitem{dwo} J. Dworkin {\it et al.}, Phys. Rev. D {\bf 41} 780 (1990).

%%Their normalization convention differs from ours in that
%each spinor has the normalization $ \sqrt{ E + m } $ factored out into their
%phase space term.  Also, to make their phase
% space correct to second order,
%one must replace $ M_{b} / M_{B} $ with $ ( E_{b} + M_{b} ) / 2M_{B} $.  

%\bibitem{bj} We employ the metric and $ \gamma $ matrix conventions of Ref.
%\cite{gar}, so that $g_1/f_1$ is { \bf positive}
% for neutron beta decay.  They differ from those of
%J.\ D.\ Bjorken and S.\ D.\ Drell,{\em Relativistic Quantum Mechanics}, 
%(McGraw-Hill, New York, 1964) only in that $ \gamma_{5} $ 
%has the opposite sign, and that there is no $ i $ in the
%definition of $ \sigma_{\alpha \beta} $.
%
%
%\bibitem{pdg} Particle Data Group, C. Caso {\it et al.}, Eur.
%Phys. J. C {\bf 3} 1, (1998).

%\bibitem{gar} A.\ Garcia and P.\ Kielanowski,
%{\em The Beta Decay of Hyperons}, Lecture Notes in Physics Vol. 222 
%(Springer-Verlag, Berlin, 1985).  With these  metric and $ \gamma $ matrix
%conventions, $g_1/f_1$ is { \bf positive}
%for neutron beta decay. 

\end{thebibliography}
\end{document}